\documentclass[a4paper,12pt]{article}
\usepackage[english]{babel}
\usepackage{jheppub2}
\usepackage{subfig}
\usepackage[T1]{fontenc} 
\usepackage{graphicx}
\usepackage{epsfig}
\usepackage{amsmath}
\usepackage{amssymb}
\usepackage{float}
\usepackage{placeins}
\usepackage{braket}
\usepackage{slashed}
\usepackage{mathdots}
\usepackage{lipsum}
\usepackage{feynmf}
\usepackage{bbm}
\usepackage{color}

\allowdisplaybreaks[4]

\title{One-loop QCD helicity amplitudes for $pp\to t\tb j$ to $O(\eps^2)$}

\author[a]{Simon Badger}
\author[a]{Matteo Becchetti}
\author[a]{Ekta Chaubey}
\author[b]{Robin Marzucca}
\author[a]{Francesco Sarandrea}
\affiliation[a]{Physics Department, Torino University and INFN Torino, Via Pietro Giuria 1, I-10125 Torino, Italy}
\affiliation[b]{Niels Bohr Institute, Copenhagen University, Blegdamsvej 17, 2100 Copenhagen \O , Denmark}
\emailAdd{simondavid.badger@unito.it}
\emailAdd{matteo.becchetti@unito.it}
\emailAdd{ekta@to.infn.it}
\emailAdd{robin.marzucca@nbi.ku.dk}
\emailAdd{francesco.sarandrea@unito.it}

\abstract{
We compute helicity amplitudes for the one-loop QCD corrections to top-quark pair production analytically in terms of a set of uniformly transcendental master integrals. We provide corrections up to $O(\eps^2)$ in the dimensional regulator for the first time which are relevant at
NNLO. Four independent pentagon integral topologies appear in the complete description
of the colour structure for which we provide numerical solutions using canonical form
differential equations and the method of generalised power series expansions. Analytic forms of the boundary values
are obtained in all cases except one where we find a one-dimensional integral representation.
}

\DeclareMathOperator{\Tr}{Tr}
\newcommand{\beq}{\begin{equation}}

\newcommand{\eeq}{\end{equation}}
\newcommand{\nn}{\nonumber}
\newcommand{\bea}{\begin{eqnarray}}
\newcommand{\eea}{\end{eqnarray}}
\newcommand{\bfig}{\begin{figure}}
\newcommand{\efig}{\end{figure}}
\newcommand{\bc}{\begin{center}}
\newcommand{\ec}{\end{center}}

\newcommand{\eps}{{\varepsilon}}
\newcommand{\tb}{{\bar{t}}}
\newcommand{\qb}{{\bar{q}}}
\newcommand{\cI}{{\mathcal{I}}}

\def\la{\langle}
\def\ra{\rangle}
\def\spA#1#2{\la#1#2\ra}
\def\spB#1#2{[#1#2]}
\def\spAB#1#2#3{\la#1|#2|#3]}

\def\spAA#1#2#3{\la#1|#2|#3\ra}
\def\spBB#1#2#3{[#1|#2|#3]}

\definecolor{mypink}{RGB}{219, 48, 122}
\definecolor{mygreen}{rgb}{0,0.7,0}
\definecolor{raspberry}{rgb}{0.53,0.15,0.34}

\date{}
\begin{document}
\maketitle
\flushbottom

\section{Introduction}

Precision predictions for the production of a pair of top-quarks in association
with a jet in hadron collisions is a high priority for current and future
experimental measurements. Next-to-next-to-leading order (NNLO) corrections in
Quantum-Chromodynamics (QCD) would allow percent level predictions for a wide
variety of observables. The theoretical challenge and the degree of calculation
complexity for such predictions remains extremely high.

Next-to-leading order (NLO) QCD corrections to $pp\to t\tb j$ were first
computed by Dittmaier, Uwer and
Weinzierl~\cite{Dittmaier:2007wz,Dittmaier:2008uj} where the amplitude level ingredients were obtained analytically. This computation was
performed using an on-shell approximation for the top quarks, corrections
including decays in the narrow width approximation~\cite{Melnikov:2010iu} and
with complete off-shell effects~\cite{Bevilacqua:2015qha} were later included
using modern numerical techniques~\cite{Ossola:2006us,Giele:2008ve,Berger:2008sj,Ellis:2008ir,Bevilacqua:2011xh,Cullen:2011ac,Cascioli:2011va}. Predictions for top quark pair production in
association with multiple jets~\cite{Hoche:2016elu} or matched to a parton shower~\cite{Alioli:2011as,Hoeche:2014qda,Czakon:2015cla} have been made possible thanks to the latest generation of automated numerical tools. The $pp\to t\tb j$ process is of particular interest since it is extremely sensitive to the top quark mass~\cite{Alioli:2013mxa,Bevilacqua:2017ipv}.

Precision predictions at NNLO are currently only available for the four
particle process $pp\to t\tb$. Advanced techniques for the subtraction of
infrared divergences~\cite{Czakon:2010td} have enabled a comprehensive range of phenomenological
studies~\cite{Czakon:2013goa,Behring:2019iiv,Catani:2019hip}. The amplitude level ingredients for these predictions are largely
known analytically~\cite{Bonciani:2008az,Bonciani:2009nb,Bonciani:2010mn,Bonciani:2013ywa,vonManteuffel:2013uoa,DiVita:2018nnh,Mastrolia:2017pfy,Becchetti:2019tjy,Badger:2021owl} although there are still a small number of non-planar double-virtual contributions that are only known numerically. 

In this article we present one previously missing ingredient relevant for a
next-to-next-to-leading computation of $pp\to t\tb j$: the expansion of the
one-loop helicity amplitudes up to $\mathcal{O}(\eps^2)$ in the dimensional
regulator. This requires the computation of new pentagon integrals that first
appear at $\mathcal{O}(\eps)$ and are one of the new results presented here.

Helicity amplitudes (including decay information for the top-quark pair in
the narrow width approximation) for this process have not been presented
analytically before. One-loop expressions for $pp\to t\tb$ production were computed in this formalism using unitarity based methods and led to relatively compact expressions~\cite{Badger:2011yu}. The motivation to do so here for the high multiplicity process is to get a sense of the
complexity that might arise in an analytic two-loop computation of $pp\to t\tb
j$. The new loop integrals appearing at $\mathcal{O(\eps)}$ depend on genuine
five-point kinematics for the first time. While at one-loop all the special
functions are of a polylogarithmic form, the alphabet is quite complex and
efficient evaluation and analytic continuation to physical kinematics is
challenging. In this context recent progress has been made to compute
analytically five-point one-loop integrals with massive external legs and
internal propagators \cite{Syrrakos:2021nij}. In this article we explore the
technique of generalised series expansions \cite{Francesco:2019yqt}, as implemented in the software \textsc{DiffExp} \cite{Hidding:2020ytt}, to numerically solve the differential equations for the master integrals. Such a technique would be applicable to two-loop integrals even in
the presence of non-polylogarithmic forms, and it has been exploited recently for several processes \cite{Abreu:2020jxa,Becchetti:2020wof,Bonciani:2021zzf,Armadillo:2022bgm}. This approach as well as related methods for the numerical solution of differential equations for master integrals~\cite{Lee:2017qql,Mandal:2018cdj,Liu:2021wks,Liu:2022chg,Liu:2022tji} have been of particular interest recently due to their wide range of applicability.

The rational coefficients of the special functions also represent a step up in
analytic complexity in comparison with previously considered two-loop five-point
amplitudes~\cite{%
Gehrmann:2015bfy,Badger:2018enw,Abreu:2018aqd,Chicherin:2018yne,Chicherin:2019xeg,Abreu:2019rpt,%
Abreu:2018zmy,Abreu:2019odu,Badger:2019djh,Abreu:2020cwb,Chawdhry:2020for,Caron-Huot:2020vlo,DeLaurentis:2020qle,Agarwal:2021grm,Abreu:2021oya,Agarwal:2021vdh,Chawdhry:2021mkw,%
Badger:2021nhg,Badger:2021imn,Abreu:2021asb,Badger:2021ega,Badger:2022ncb}. We present a complete set of partial colour amplitudes in terms of
master integrals valid to all orders in the dimensional regulator. These objects
are considerably more complex than the four-dimensional limits and we employ a
finite field reconstruction technique~\cite{Peraro:2016wsq,Peraro:2019svx} and a rational parametrisation of the
kinematics based on momentum twistors~\cite{Hodges:2009hk} to overcome the algebraic complexity.

Our paper is organised as follows. We begin by reviewing the colour
decomposition of the amplitudes in both $t\tb ggg$ and $t\tb q\qb g$ channels
and describe the infrared and ultraviolet pole structure. We then describe the
finite field reconstruction approach taken to extract the independent helicity
amplitudes.  We then turn our attention to the evaluation of the master
integrals. We present canonical form differential equations for the four
independent topologies appearing in our process. The computation of the
boundary terms is described and the numerical evaluation using generalised
series expansions with DiffExp is presented. Finally we present some numerical
results before giving an outlook for the future.

\section{Colour decomposition and infrared pole structure \label{sec:cdecomp}}

We choose to define the two partonic channels for $pp\to t \tb j$ with all
momenta out-going. Evaluation for physical kinematics can be performed using
the appropriate analytic continuation. We write the amplitudes according to the colour decomposition~\cite{Bern:1994fz}. Therefore, for the process $0\to
\tb t g g g$ we have: 
\begin{align}
  \mathcal{A}^{(L)}&(1_\tb, 2_t, 3_g, 4_g, 5_g) = g_s^{3+2L} N_\eps^L \bigg\{ \nonumber\\
      &\sum_{\sigma\in S_3} (t^{a_{\sigma(3)}} t^{a_{\sigma(4)}} t^{a_{\sigma(5)}} )_{i_2}^{\bar{i}_1} A^{(L)}_1(1_\tb, 2_t, \sigma(3)_g, \sigma(4)_g, \sigma(5)_g) \nonumber\\
    + &\sum_{\sigma\in S_3/\mathbbm{Z}_{2}} \delta^{a_{\sigma(3)} a_{\sigma(4)}} (t^{a_{\sigma(5)}} )_{i_2}^{\bar{i}_1} A^{(L)}_2(1_\tb, 2_t, \sigma(3)_g, \sigma(4)_g, \sigma(5)_g) \nonumber\\
    + &\sum_{\sigma\in S_3/\mathbbm{Z}_{3}} \Tr(t^{a_{\sigma(3)}} t^{a_{\sigma(4)}} t^{a_{\sigma(5)}}) \delta_{i_2}^{\bar{i}_1} A^{(L)}_3(1_\tb, 2_t, \sigma(3)_g, \sigma(4)_g, \sigma(5)_g)
    \bigg\}.
  \label{eq:colourdecomp_2t3g}
\end{align}
Here we have used $g_s$ to denote strong coupling and taken an overall normalisation 
\begin{equation}
  N_\eps = \frac{e^{\eps\gamma_E}\Gamma^2(1-\eps)\Gamma(1+\eps)}{(4\pi)^{2-\eps}\Gamma(1-2\eps)} \,.
  \label{eq:ampnorm}
\end{equation}
$A^{(L)}_i$ are the partial amplitudes which appear in the full amplitude as
sums over permutations of the momenta. $S_3$ indicates the six permutations of
the three gluons while $S_3/\mathbbm{Z}_{2}$ and $S_3/\mathbbm{Z}_{3}$ are smaller symmetry groups with 3 and 2 elements respectively.
The $SU(N_c)$ colour structures are written using the fundamental generators
$(t^a)_i^{\bar{j}}$ where $a=1,\cdots,8$ are indices of the adjoint representation, while $i,\, \bar{j}=1,2,3$ are indices in the fundamental and anti-fundamental representation respectively.

Following the same conventions, we colour decompose the process $0\to \tb t \qb q g$ as (see for example~\cite{Kunszt:1994nq}),
\begin{align}
  \mathcal{A}^{(L)}&(1_\tb, 2_t, 3_q, 4_\qb, 5_g) = g_s^{3+2L} N_\eps^L \bigg\{ \nonumber\\
	    &\delta^{\bar{i}_4}_{i_1}(t^{a_{5}})^{\bar{i}_2}_{i_3} A^{(L)}_{1}(1_\tb, 2_t, 3_\qb, 4_q, 5_g)\nonumber\\
    + &\delta^{\bar{i}_3}_{i_2}(t^{a_{5}})^{\bar{i}_4}_{i_1} A^{(L)}_{2}(1_\tb, 2_t, 3_\qb, 4_q, 5_g)\nonumber\\
    - &\frac{1}{N_c} \delta^{\bar{i}_2}_{i_1}(t^{a_{5}})^{\bar{i}_4}_{i_3} A^{(L)}_{3}(1_\tb, 2_t, 3_\qb, 4_q, 5_g)\nonumber\\
    - &\frac{1}{N_c} \delta^{\bar{i}_4}_{i_3}(t^{a_{5}})^{\bar{i}_2}_{i_1} A^{(L)}_{4}(1_\tb, 2_t, 3_\qb, 4_q, 5_g)
    \bigg\}.
  \label{eq:colourdecomp_2t2q1g}
\end{align}
Each of the partial amplitudes is further decomposed into a polynomial in $N_c$
and the number of light and heavy flavours, $n_f$ and $n_h = 1$ respectively.
Suppressing the momentum arguments we have
\begin{align}
  A^{(0)}_1 &= A^{(0)}_{1;0} = A^{(0)} \\
  A^{(0)}_2 &= 0 \\
  A^{(0)}_3 &= 0 \\
  A^{(1)}_1 &= N_c A^{(1)}_{1;1} + \frac{1}{N_c} A^{(1)}_{1;-1} + n_f A^{(1),f}_{1;0} + A^{(1),h}_{1;0} \\
  A^{(1)}_2 &= A^{(1)}_{2;0} \\
  A^{(1)}_3 &= A^{(1)}_{3;0}
\end{align}
for the $0\to \tb t g g g$ channel, and
\begin{align}
  A^{(0)}_X &= A^{(0)}_{X;0} \\
  A^{(1)}_X &= N_c A^{(0)}_{X;1} + \frac{1}{N_c} A^{(1)}_{X;-1} + n_f A^{(1),f}_{X;0} + A^{(1),h}_{X;0}
\end{align}
for $0\to \tb t q \qb g$ channel where $X=1,\cdots,4$.

The kinematics for these processes is:
\begin{equation} \label{eq:kinematics}
 p_1^2 = p_2^2 = m_t^2, \,\,\, p_3^2 = p_4^2 = p_5^2 = 0, \,\,\, d_{ij} = p_i \cdot p_j , \,\,\, s_{ij} = (p_i + p_j)^2 
\end{equation}
where $p_1$ and $p_2$ are the momenta of the external top quarks, $p_3$ and $p_4$ are the momenta associated either to the a pair of gluons or a pair of massless quarks, and $p_5$ is the momentum of the remaining gluon. All the particles are on-shell and the top quarks are considered to be massive, with $m_t$ the top mass. Finally, throughout this paper, we work both with the kinematic invariants $d_{ij}$ and $s_{ij}$ as defined in \eqref{eq:kinematics}. \par


\subsection{Infrared Singularities \label{ssec:irpoles}}

Catani, Dittmaier and Trocsanyi (CDT) were the first to present a closed formula for
the universal infrared (and ultraviolet) pole structure of an arbitrary
one-loop amplitude with massless and massive QCD partons~\cite{catani2001one}.
Using the colour space notation~\cite{Catani:1996vz} the factorisation of the infrared poles can be denoted simply as,
\begin{align}
	\ket{\mathcal{A}^{(1)}_{n}} = \mathbf{I}_{n}\ket{\mathcal{A}^{(0)}_{n} } + \ket{\mathcal{A}^{(1)}_{n}}^{\mbox{fin}}+ O(\eps).
	\label{eq:irdecomp}
\end{align}
For renormalised amplitudes the pole operator $\mathbf{I}_{n}$ is defined as\footnote{We omit the imaginary parts in our reproduction of the  $\mathbf{I}_{n}$ operator. For a correct treatment across the full physical phase-space the prescription is given in Ref.~\cite{catani2001one}. At the test points we provide, the form given and \textsc{Mathematica}'s internal prescription are sufficient to find agreement up to $\mathcal{O}(\eps^{-1})$.},
\begin{align} \label{eq:CDTpoles}
	\mathbf{I}_n = N_\eps \left(
    \sum_{i,j=1}^{n} \mathbf{T}_{i} \cdot \mathbf{T}_{j} \left(\frac{\mu_R^2}{-2 d_{ij}} \right)^{\eps} \mathcal{V}_{ij}
  - \sum_{j=1}^{n} \Gamma_{j} \right)
\end{align}
where we have followed the normalisation conventions from Eq.
\eqref{eq:ampnorm}\footnote{Note that the difference between $N_\eps$ and the
factor used in reference~\cite{catani2001one} appears at $\mathcal{O}(\eps^3)$
and therefore does not effect the one-loop singluarities.}. The function $\mathcal{V}_{ij}$
arises from the soft singularities which contains colour correlations of the
form $\mathbf{T}_{i} \cdot \mathbf{T}_{j}$,
\begin{align}
  \mathcal{V}_{ij} =
  \begin{cases}
    \dfrac{1}{\eps^2} & \text{$i$ and $j$ are massless} \\
    \dfrac{1}{2\eps^2}
    + \dfrac{1}{2\eps}\log\left( \dfrac{m_j^2}{-2 d_{ij}} \right)
    - \frac{1}{4}\log^2\left( \dfrac{m_j^2}{-2 d_{ij}} \right) - \dfrac{\pi^2}{12}
    & \text{$i$ massless, $j$ massive} \\
    \begin{aligned}
      &\frac{2 d_{ij}}{(s_{ij}-(m_i-m_j)^2) \beta_{ij} \eps} \log\left( -\frac{1+\beta_{ij}}{1-\beta_{ij}} \right) \\
      - &\frac{1}{4}\left( \log^2\left( \frac{m_i^2}{-2 d_{ij}}\right) + \log^2\left( \frac{m_j^2}{-2 d_{ij}} \right)\right) - \frac{\pi^2}{6}
    \end{aligned}
    & \text{$i$ and $j$ are massive}
  \end{cases}
\end{align}
where $\beta_{ij} = \sqrt{1-\tfrac{4 m_i m_j}{s_{ij}-(m_i-m_j)^2}}$ comes
from the kinematic threshold for the production of a top quark pair. Since we
only have two massive partons in our process with the same mass we have a
single threshold $\beta_{12} = \beta(s_{12},m_t^2) = \sqrt{1-\frac{4m_t^2}{s_{12}}}$. The
finite parts of the function $\mathcal{V}_{ij}$ will not play a role in the
cross checks of our computation though they are important to ensure the correct
small mass limits, as stated in~\cite{catani2001one}. The functions $\Gamma_{j}$ arise
from the hard collinear region and depend on the anomalous dimensions of the partons. Our amplitudes are computed including
wave-function renormalisation but excluding coupling renormalisation and
therefore additional UV poles proportional to the QCD $\beta$
function are present in the expressions.

We will compute all partial amplitudes in terms of master integrals valid to
all orders in $\eps$. The verification of the infrared pole structure is
therefore an extremely strong check on the validity of our expressions. The
inclusion of the wave-function renormalisation counter-terms ensures the amplitude is
gauge invariant which also provides a strong cross check.

Explicit evaluations of the CDT formula, including for the two partonic channels, into
the partial decompositions Eq.~\eqref{eq:colourdecomp_2t3g} and Eq.~\eqref{eq:colourdecomp_2t2q1g} are given in Appendix~\ref{app:irpoles}.

\section{Helicity Amplitude Setup \label{sec:helampcomp}}

In this section we describe the computational set up for the helicity
amplitudes. This makes use of the well known spinor-helicity formalism for
massless and massive particles.

\subsection{Helicity amplitudes \label{ssec:helamp}}

The helicity states for the massive fermions are computed using the standard decomposition along an arbitrary reference direction~\cite{Kleiss:1985yh}:
\begin{align}
  u_+(p,m) = \frac{(\slashed{p}+m)|n\ra}{\spA{p^\flat}{n}}
\end{align}
where $p$ is the massive fermion momentum, $m$ is the mass, $n$ is the
arbitrary reference direction and $p^\flat = p - \tfrac{m^2}{2p\cdot n} n$.
Since the direction $n$ is arbitrary the positive helicity state is related to
the negative helicity state through the transformation $n\leftrightarrow
p^\flat$ together with a normalisation factor accounting for the change in
spinor phase. For further details of this relation, and other aspects of the
massive spinor-helicity formalism used in this article, we point the reader to~\cite{Badger:2021owl}
and references therein.

We perform an analytic reconstruction in the minimal set of six
on-shell variables making use of the following basis for spin structures,
\begin{align} \label{eq:spindecomp}
  A_x^{(L)}(1_t^+,2_\tb^+,3^{h_3}&,4^{h_4},5^{h_5};n_1,n_2) =
  m_t \Phi(3^{h_3},4^{h_4},5^{h_5})\nonumber\\&
  \sum_{i=1}^4 \Theta_i(1,2;n_1,n_2) A_x^{(L),[i]}(1_t^+,2_\tb^+,3^{h_3},4^{h_4},5^{h_5}).
\end{align}
The decomposition involves a phase factor $\Phi$ to account for the massless
parton helicities, four basis functions $\Theta_i$ for the spin dependence of
the top-quark pair and the associated subamplitudes $A_x^{(L),[i]}$. This
representation and notation has been introduced in the recent study of
top-quark production~\cite{Badger:2021owl}. The functions $\Theta$ contain all
dependence on the arbitrary reference vectors introduced to define the positive
helicity massive fermions. Such a spin basis is not unique and the
normalisations for the $\Theta$ functions have been chosen such that all
subamplitudes have the same dimension and are free of any spinor phase. This
form is sufficient to account for the decays of the top quarks in the narrow
width approximation~\cite{Melnikov:2010iu,Campbell:2012uf}.

For the amplitudes considered in this article the explicit forms for $\Phi$ and $\Theta$ are:
\begin{align}
  \Phi(3^+,4^+,5^+) &= \frac{\spB35}{\spA34\spA45}, \\
  \Phi(3^+,4^+,5^-) &= \frac{\spAA{5}{p_3 p_4}{5}}{\spA34^2}, \\
  \Phi(3^+,4^-,5^+) &= \frac{\spAA{4}{p_5 p_3}{4}}{\spA35^2}
  \label{eq:gluonphases}
\end{align}
and,
\begin{align}
  \Theta_1(1,2,n_1,n_2) &= \frac{\spA{n_1}{n_2}s_{34}}{\spA{1^\flat}{n_1}\spA{2^\flat}{n_2}}, \\
  \Theta_2(1,2,n_1,n_2) &= \frac{\spA{n_1}{3}\spA{n_2}{4}\spB34}{\spA{1^\flat}{n_1}\spA{2^\flat}{n_2}}, \\
  \Theta_3(1,2,n_1,n_2) &= \frac{\spA{n_1}{3}\spA{n_2}{3}\spBB{3}{p_4p_5}{3}}{s_{34}\spA{1^\flat}{n_1}\spA{2^\flat}{n_2}}, \\
  \Theta_4(1,2,n_1,n_2) &= \frac{\spA{n_1}{4}\spA{n_2}{4}\spBB{4}{p_5p_3}{4}}{s_{34}\spA{1^\flat}{n_1}\spA{2^\flat}{n_2}}.
  \label{eq:spinbasis}
\end{align}
The amplitudes with a massless fermion pair use the same spin decomposition \eqref{eq:spinbasis} but with different phases:
\begin{align}
  \Phi(3_q^-,4_\qb^+,5^+) &= \frac{\spA34}{\spA45^2}, \\
  \Phi(3_q^+,4_\qb^-,5^+) &= \frac{\spA34}{\spA35^2}.
  \label{eq:quarkphases}
\end{align}

\subsection{Rational phase space parametrisation \label{ssec:momtwist}}

Our computation uses a rational phase-space parametrisation together with a
numerical sampling of the relevant set of Feynman diagrams using modular
arithmetic. The generation of this rational parametrisation uses the momentum twistor
formalism~\cite{Hodges:2009hk}, a technique that has been applied numerous times in
similar amplitude computations. For the case of a top-quark pair plus three
massless partons the method is essentially the same as the one described in
Ref.~\cite{Badger:2021owl} for a top-quark pair plus two massless partons.

We begin by generating a rational parametrisation for configurations of seven massless particles (11 free variables) with momenta $q_1,\cdots q_7$\footnote{The
specific form of the massless configuration is not important. A few all
multiplicity parametrisations have been presented in the literature
\cite{Badger:2016uuq, Buciuni:Thesis, Pogel:2021buy}.}. The five particle system for
$t\bar{t}$ plus three partons can then be written,
\begin{align}
  p_1 &= q_1+q_2,& 
  p_2 &= q_3+q_4,&
  p_3 &= q_5, &
  p_4 &= q_6, &
  p_5 &= q_7,
\end{align}
with additional constraints to ensure the massive momenta $p_1$ and $p_2$ are on-shell. Specifically we solve the constraints:
\begin{align}
  q_1 \cdot q_2 &= q_3 \cdot q_4, &
  \spA{q_2}{q_5} &= 0, &
  \spB{q_2}{q_5} &= 0, &
  \spA{q_4}{q_5} &= 0, &
  \spB{q_4}{q_5} &= 0.
\end{align}
Having found the rational parametrisation we change variables to,
\begin{align}
  s_{34} &= (p_3+p_4)^2, \\
  t_{12} &= s_{12}/s_{34}, \\
  t_{23} &= (s_{23}-m_t^2)/s_{34}, \\
  t_{45} &= s_{45}/s_{34}, \\
  t_{15} &= (s_{15}-m_t^2)/s_{34}, \\
  x_{5123} &= -\frac{\spAA{5}{p_1 p_{45}}{3}}{\spA53 s_{12}}.
\end{align}
In the last variable we have introduced the notation $p_{ij} = p_i+p_j$. We note that the only dimensionful variable $s_{34}$ can be set to $1$ and
restored easily through dimensional analysis. It is not possible to use the top
quark mass as a variable without introducing square roots and hence we choose a
spinorial trace. For completeness we present explicitly the map from Lorentz invariants to the rational parametrisation:
\begin{align}
  d_{12} &= \frac{s_{34}t_{12}}{2 t_{45}}\Bigg( t_{45} + 2 t_{45} (-1 + t_{51}) x_{5123} \nonumber\\& + 2 t_{12} t_{45} x_{5123}^2 - 2 (t_{51} + (-1 + t_{12}) x_{5123}) (t_{23} + t_{12} x_{5123}) \Bigg), \\
  d_{23} &= \frac{s_{34}t_{23}}{2}, \\
  d_{34} &= \frac{s_{34}}{2}, \\
  d_{45} &= \frac{s_{34}t_{45}}{2}, \\
  d_{15} &= \frac{s_{34}t_{51}}{2}, \\
  m_t^2 &= \frac{s_{34}t_{12}}{t_{45}}\Bigg( t_{23} (t_{51} + (-1 + t_{12}) x_{5123}) \nonumber\\&+ x_{5123} (t_{45} + t_{12} t_{51} - t_{45} t_{51} + t_{12} (-1 + t_{12} - t_{45}) x_{5123}) \Bigg).
\end{align}

\subsection{Reduction to master integrals \label{ssec:reduction}}

\begin{figure}[h]
  \begin{center}
    \includegraphics[width=5cm]{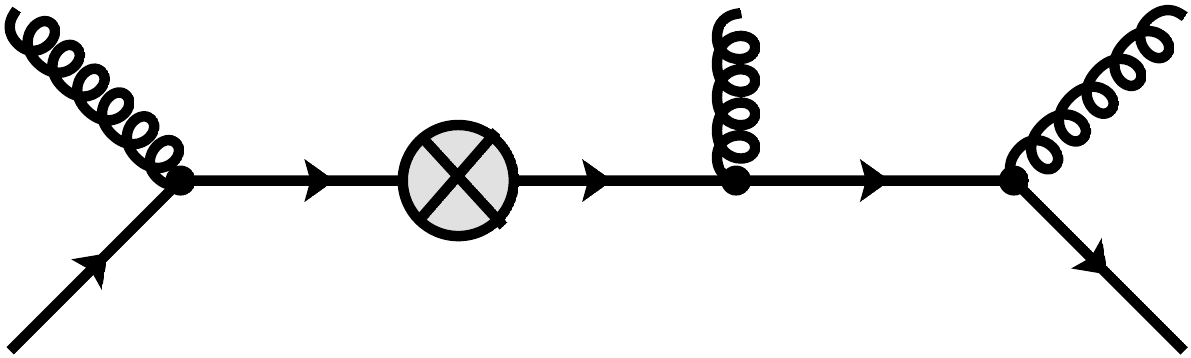}
  \end{center}
  \caption{Example renormalisation counter-term diagram contributing to $t\bar{t}+3g$ one-loop amplitude.}
  \label{fig:ct}
\end{figure}

Our amplitude computation strategy follows the method applied to recent
two-loop computations~\cite{Badger:2021owl}. The colour-ordered helicity amplitudes are first
generated from Feynman diagrams using \textsc{QGraf}~\cite{Nogueira:1991ex} which are then
processed using a combination of \textsc{Mathematica} and \textsc{FORM}~\cite{Kuipers:2012rf,Ruijl:2017dtg}
scripts. The \textsc{Spinney} package is used to process parts of the numerator
algebra~\cite{Cullen:2010jv}. The integral topologies are identified from the
loop momentum dependent propagators and the coefficients of the loop-dependent
numerators are computed using the momentum twistor parametrisation. A tensor
integral representation for each diagram is obtained using a basis of
irreducible numerators for the maximal cut topologies. Since the maximal cut
has four independent momenta, no transverse
integration~\cite{Mastrolia:2016dhn} step is required in contrast to the method
of reference~\cite{Badger:2021owl}. The diagram numerators are computed using a
symbolic value for the spin dimension $d_s=g^\mu{}_\mu$ and our results are
presented as an expansion around $d_s=2$,
\begin{align}
  A_x^{(L),[i]} = A_x^{(L,0),[i]} + (d_s-2) A_x^{(L,1),[i]}.
\end{align}
The dimension $d_s$ used in the numerator algebra is kept separate from the
loop integration dimension $d=4-2\eps$ in order to account for the scheme
dependence between the FDH ($d_s=4$) and tHV ($d_s=4-2\eps$) schemes.
The diagram numerators are then reduced to master integrals via
integration-by-parts identities~\cite{Tkachov:1981wb,Chetyrkin:1981qh} following the Laporta algorithm~\cite{Laporta:2001dd}. The
complete reduction is implemented in \textsc{LiteRed}~\cite{Lee:2012cn} and \textsc{FiniteFlow}~\cite{Peraro:2019svx} so that
numerical sampling with modular arithmetic can be used to reconstruct the
coefficients of the master integrals directly without analytic intermediate
steps. Wave-function renormalisation must also be performed in order to obtain
a gauge invariant result. We generate diagrams with the counter-term insertions
which are added to the one-loop numerators as shown in figure \ref{fig:ct}. The
counter-terms are written in terms of loop integrals valid to all orders in
$\eps$. The Feynman rule for the counter-term insertion can be written as,
\begin{align}
  \raisebox{-1mm}{\includegraphics[width=2cm]{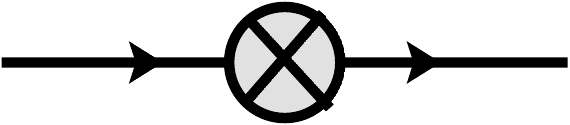}}
  =
  \raisebox{-0.5mm}{\includegraphics[width=1.5cm]{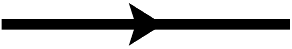}}
  \left( N_c - \frac{1}{N_c} \right)\left( 1 + \frac{(d_s-2)(1-2\eps)}{4(1-\eps)} \right)
  \raisebox{-3mm}{\includegraphics[width=2cm]{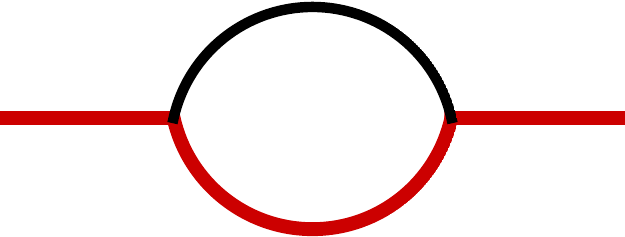}}.
  \label{eq:ct}
\end{align}
The right hand side of this equation consists of: a line representing the Feynman rule for a massive fermion propagator, a rational function of $N_c, d_s $ and $\eps$, and a wave-function bubble graph representing the Feynman integral with one massive and one massless propagator and a mass scale of $m_t^2$. We note that using this integral form for the counter-term, the amplitude is already guage invariant at the level of master integrals. This ensures that only gauge invariant quantities are reconstructed analytically and allows us to sidestep the issue of including external leg corrections with on-shell ingredients~\cite{Ellis:2008ir,Britto:2011cr,Badger:2017gta}.

The coefficients of the master integrals are functions of the dimensional
regularisation parameter $\eps$ and the six free parameters in the rational
phase-space. Since this is a one-loop problem the evaluation time for the
amplitude within the \textsc{FiniteFlow} setup is quite fast and so functional
reconstruction of high degree polynomials is feasible. Nevertheless, we find
that rationalising the phase-space increases the polynomial degree
significantly so in addition we apply linear relations and a univariate partial
fractioning to the master integral coefficients before reconstruction as has been effective in many cases with massless propagators(for example~\cite{Abreu:2018zmy}). We apply
the univariate reconstruction method and the algorithm for linear relations described in Ref.~\cite{Badger:2021imn}. The
first step in this method requires a matching of denominator (and numerator)
factors for which we build an ansatz from a set of spinor products, Gram
determinants and other denominators appearing in the differential equations which
we describe in the next section. We have used the following kinematic structures to generate our factor ansatz,
\begin{align}
  \{&
  \eps, 1-\eps, 1-2\eps, 3-2\eps,\nonumber\\&
  \spA{3}{4}, \spB{3}{4}, \spAB{3}{1}{4},\nonumber\\&
  d_{12}, d_{12}+m_t^2, d_{12}-m_t^2, d_{13},\nonumber\\&
  s_{12}, s_{13}, s_{34}, s_{12}-s_{34}, s_{13}-s_{24},\nonumber\\&
  (p_{23}\cdot p_1)^2 - m_t^2 s_{23} = \Delta_3(p_{23},p_1)^2,\nonumber\\&
  \spAA{3}{p_1p_{12}}{4}, \spBB{3}{p_1p_{12}}{4}, \spAA{3}{p_1p_2}{4}, \spBB{3}{p_1p_2}{4}, \spAA{3}{p_1p_2}{3}, \spBB{3}{p_1p_2}{3},\nonumber\\&
  \spAA{3}{p_2p_5p_3p_1}{4}+m_t^2 s_{35} \spA34,
  \spBB{3}{p_2p_5p_3p_1}{4}+m_t^2 s_{35} \spB34,\nonumber\\&
  {\rm tr}_5(3451) = \spAB{3}{p_1p_5p_4}{3}-\spAB{3}{p_4p_5p_1}{3},\nonumber\\&
  (d_{13} d_{25} - p_3\cdot p_{24} p_4\cdot p_{13})\spAA{3}{p_1p_{12}}{4} + 2 \spA34 d_{13} d_{24} p_5\cdot p_{34},\nonumber\\&
  (d_{13} d_{25} - p_3\cdot p_{24} p_4\cdot p_{13})\spBB{3}{p_1p_{12}}{4} + 2 \spB34 d_{13} d_{24} p_5\cdot p_{34},\nonumber\\&
  |Y_5|
  \} \,,
  \label{eq:coeffansatz}
\end{align}
where we have introduced a notation for the 3-mass triangle Gram determinants, $\Delta_3$ and the Cayley matrix associated with the pentagon integral with four internal masses,
\begin{align}
  (Y_5)_{ij} = -p_{i,j-1}^2 + m_i^2 + m_j^2 \,,
\end{align}
where $p_{i,j-1} = \sum_{k=i}^{j-1} p_k$, and $m = \{m_t,0,m_t,m_t,m_t\}$.  We note that in spinor-helicity variables, many of the Gram and Cayley determinants
factorise and so we don't need to specify all Gram and Cayley determinants explicitly.

To generate an ansatz that matches all denominators this list is permuted over
an (overcomplete) set of six permutations of $3,4,5$ and two permutations of
$1,2$ ($12$ in total) and duplicate entries are removed.  To match the polynomial
factors this list is evaluated using the rational momenta parametrisations from
which a list of independent polynomials can be extracted.

We summarise our reduction strategy as follows: we apply the rational kinematic parametrisation to the processed Feynman graphs for the four different spin projections. Each of these projected amplitudes are reduced to master integrals and reconstructed using \textsc{FiniteFlow}. After reconstruction, the projected amplitudes are used to construct the sub-amplitudes, again with the help of reconstruction over finite fields, linear relations and univariate partial fractioning.

\section{Computation of the Master Integrals}

\begin{figure}[h]
  \begin{center}
    \includegraphics[width=7cm]{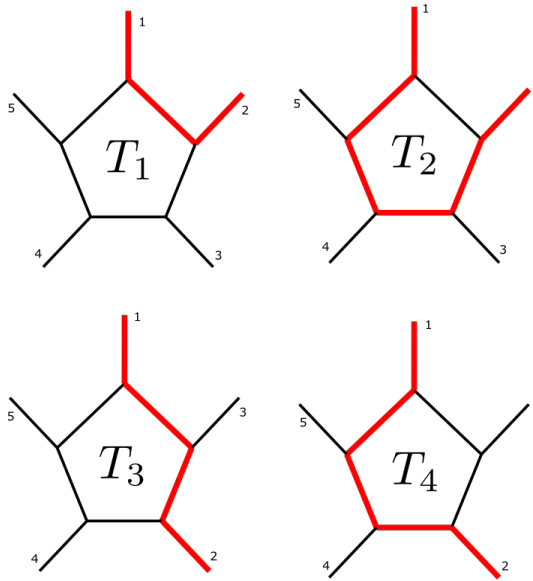}
  \end{center}
  \caption{The four distinct one-loop integral topologies appearing in the $pp\to t\tb j$ amplitudes. Black lines denote massless particles while red lines denote massive particles.}
  \label{fig:pentagontopologies}
\end{figure}

There are four distinct pentagon function topologies appearing in the
amplitudes as shown in Figure \ref{fig:pentagontopologies}. To find the minimal
set of master integrals (MIs) which describes each topology we perform
Integration-By-Parts (IBPs) reduction \cite{Chetyrkin:1981qh,Chetyrkin:1979bj},
as implemented in the software \textsc{LiteRed} \cite{Lee:2012cn,Lee:2013mka}
and \textsc{FiniteFlow} \citep{Peraro:2019svx}. We find that the topologies
$T_1$, $T_2$, $T_3$ and $T_4$ are described, respectively, by 15, 21, 17 and 19
MIs (see figures \ref{fig:T1-mis},\ref{fig:T2-mis},\ref{fig:T3-mis} and
\ref{fig:T4-mis}). By employing symmetry relations among the different
topologies, and their permutations, we find that the amplitudes can be written
in terms of minimal set of 130 MIs across the four topologies. The evaluation
of the MIs that appear in the amplitudes is discussed in section
\ref{sec:res_mis}.

We compute the MIs, $\vec{f}(x,\eps)$, by means of the differential
equations method \citep{Kotikov:1990kg,Remiddi:1997ny,Gehrmann:1999as}.
Specifically, we work with a system of differential equations in canonical form
\cite{Henn:2013pwa}:
\begin{equation} \label{eq:deqsCan}
 d \vec{f}(\vec{x},\eps) = \eps \, d \, A(\vec{x}) \vec{f}(\vec{x},\eps),
\end{equation}   
where $d$ is the total differential with respect to the kinematic invariants,
\begin{equation}
\vec{x} = \left\{d_{12}, d_{23}, d_{34}, d_{45}, d_{15}, m_t^2\right\}.
\end{equation}   
The matrix $A(\vec{x})$ is a linear combination of logarithms:
\begin{equation}
 A(\vec{x}) = \sum c_i \log (\alpha_i (\vec{x})),
\end{equation}
where $c_i$ are matrices of rational numbers and $\alpha_i (\vec{x})$ are
algebraic functions of the kinematic invariants $\vec{x}$. \par A major feature
of the systems of differential equations for the four  pentagon topologies is
that they depend on the following set of square roots:
\begin{align}
  \beta(a_1,m^2) &= \sqrt{1-\frac{4 m^2}{a_1}}, \nn \\
  \Delta_3\left(P,Q\right) &= \sqrt{(P\cdot Q)^2 - P^2 Q^2}, \nn \\
  \operatorname{tr}_5 &= \operatorname{tr}_5(3,4,5,1) = \sqrt{\operatorname{det}G(p_3,p_4,p_5,p_1)}, 
  \label{eq:sqrt}
\end{align}
where the argument $a_1$ can be functions of the kinematic invariants, $P$ and $Q$ are momenta and
$G_{ij}(\vec{v}) = 2 v_i\cdot v_j $ is the Gram matrix.

We choose to solve the systems of differential equations for the MIs using the
generalized power series expansion method \cite{Francesco:2019yqt}, as
implemented in the software \textsc{DiffExp} \cite{Hidding:2020ytt}. Although
this method furnishes a semi-analytic solution to the MIs, it has the advantage
of allowing a fast and high precision numerical evaluation. Moreover, the
analytic continuation of the solution is easier with respect to an analytic
approach. Indeed, while it could be possible to linearize the square roots
system\footnote{We checked explicitly that it is possible for topology $T_1$.}
\eqref{eq:sqrt} with a transformation of the kinematic invariants, and obtain
an analytic solution in terms of \emph{polylogarithmic functions} (MPLs)
\cite{Goncharov:1998kja,Goncharov:2001iea}, the system of differential
equations will involve polynomials of high degree in the linearized variables.
This feature impacts significantly the computation since the system of
differential equations in the new set of variables is too large to be handled
efficiently. In addition, the determination of the phase-space regions, and
therefore the analytic continuation, is more complicated.

We finish the first part of this section by discussing a few more details of
our computation. Firstly, we would like to stress that we reconstruct the
systems of differential equations for the four pentagon topologies exploiting
finite fields methods, implemented in \textsc{FiniteFlow}, for a basis of
master integrals, $\vec{f'}$ that does \textit{not} contain the square roots
\eqref{eq:sqrt}:
\begin{equation} \label{eq:deqsNoRoots}
 d \vec{f'}(\vec{x},\eps) = \, d \, A'(\vec{x},\eps) \vec{f'}(\vec{x},\eps).
\end{equation}
In doing so, we obtain a system of differential equations that is not in a
canonical form, but we also avoid dealing with square roots in the
reconstruction procedure. Then, an $\eps$-factorized form can be achieved
by a rotation of the basis of MIs, $\vec{f} = B(\vec{x})\vec{f'}$, under which
the matrix $d\, A'(\vec{x},\eps)$ transforms as: \begin{equation} d\,
  A'(\vec{x},\eps) \rightarrow B^{-1}(\vec{x}) d \, A'(\vec{x},\eps)
  B(\vec{x}) - B^{-1}(\vec{x}) d\, B(\vec{x}) = \eps \, d \, A(\vec{x}) ,
\end{equation} where $B(\vec{x})$ is a diagonal matrix whose entries are the
square roots \eqref{eq:sqrt}. Obtaining the matrix $A'(\vec{x},\eps)$ can in general be complicated for multi-loop cases but for this one-loop case it is straightforward. The rotation matrix $ B(\vec{x})$ is easily obtained using information from the maximal cuts of the topologies.  

Finally, we comment on the computation of the boundary conditions for the
system \eqref{eq:deqsCan}. We compute the boundary conditions, for a minimal
subset of the MIs, by direct integration of their Feynman parameter representation at the
kinematic point:
\begin{equation}
\vec{x}_0 :=  \left(-2, -2, -2, -2, -2, 1\right).
\end{equation}
This step is performed using the linear-reducibility strategy as implemented in
\textsc{HyperInt} \cite{Panzer:2014caa} with the help of \textsc{PolyLogTools}
\cite{Duhr:2019tlz}. High precision numerical boundary values are obtained for
the remaining integrals using $\textsc{DiffExp}$. We detail the boundary
condition computation for each topology in the following subsections.

We include ancillary files containing the systems of differential equations
\eqref{eq:deqsCan} for all the four topologies, the analytic expressions and
the numerical values of the boundary conditions, and a $\textsc{DiffExp}$
template for a standalone evaluation of the MIs.

\subsection{The $T_1$ topology with one massive internal propagator}

\begin{figure}[h]
  \begin{center}
    \includegraphics[scale=0.45]{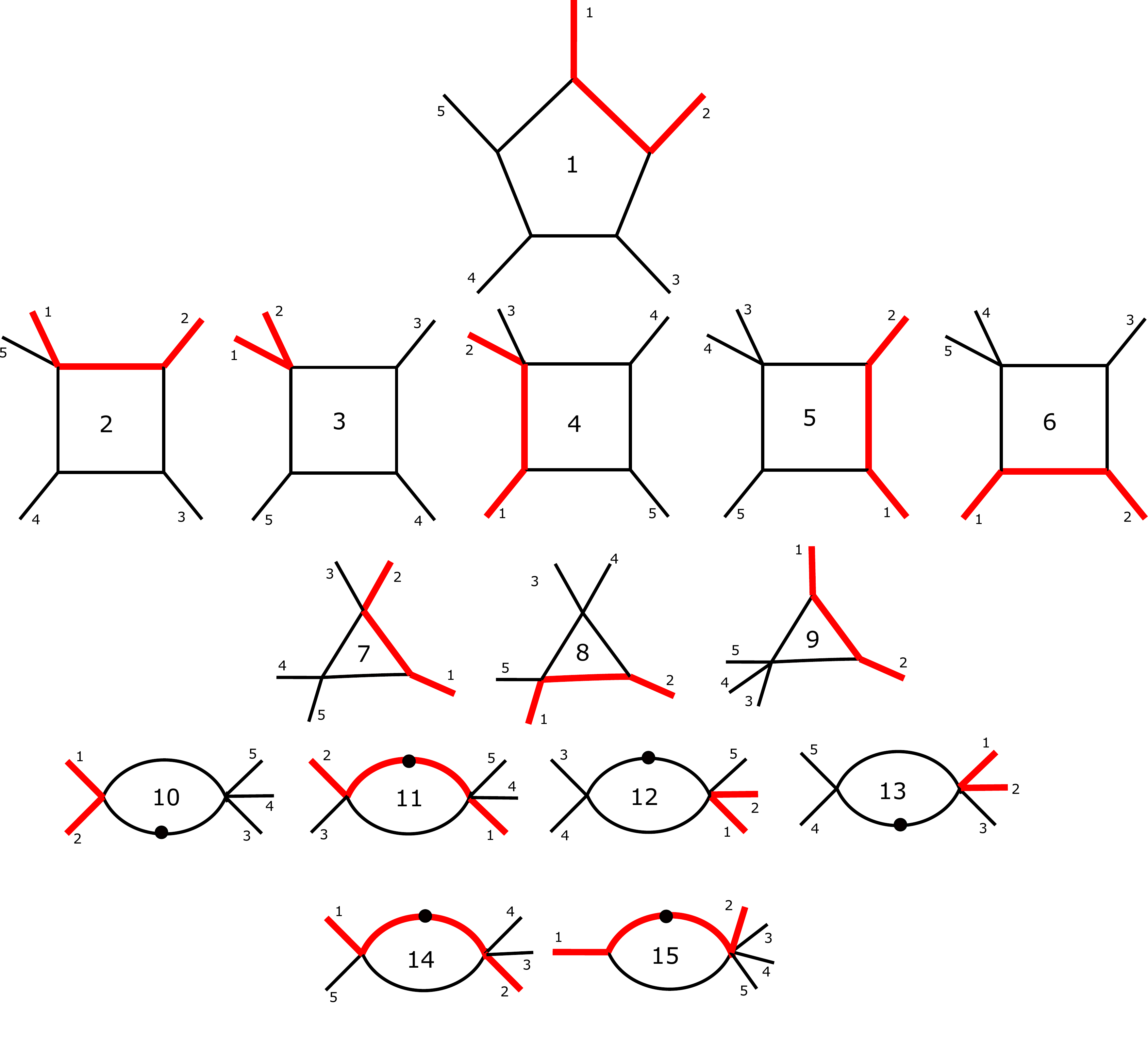}
  \end{center}
  \caption{The 15 master integrals appearing in the $T_1$ topology, denoted by $\cI$'s in~\eqref{eq:T1-utbasis}. Red/black lines indicate massive/massless particles respectively. Dotted internal lines indicate propagators with an additional power in the denominator of the integral. Each integral is associated with a kinematic normalisation which ensures the basis leads to a canonical form differential equation.}
  \label{fig:T1-mis}
\end{figure}

There are 15 master integrals in the topology $T_1$ as shown in Figure \ref{fig:T1-mis}. The integrals are defined as:
\begin{equation}
   \label{Topo}
   \mathcal{I}^{T_1,[d]}_{a_1,a_2,a_3,a_4,a_5} = \int \mathcal{D}^d k_1 \dfrac{1}{D_1^{a_1}D_2^{a_2}D_3^{a_3}D_4^{a_4}D_5^{a_5}},
\end{equation}
where
\begin{align}
  D_1 &= k_1^2, \; D_2 = (k_1-p_1)^2 -m_t^2, \; D_3 = (k_1 -p_1 -p_2)^2, \nonumber\\
  D_4 &= (k_1 + p_4 + p_5)^2, \; D_5 = (k_1 + p_5)^2,
\end{align}
$a_i$ are positive integers, $d = d_0 - 2\eps$ is the space-time dimension, and the integration measure is defined as
\begin{equation}
  \mathcal{D}^d k_1 = \dfrac{d^d k_1}{i \pi^{\frac{d}{2}}} e^{\eps \gamma_E} \left(\dfrac{m_t^2}{\mu^2}\right)^{\eps} \,.
\end{equation}
The basis of canonical MIs is chosen to be:
\begin{align}
\label{eq:T1-utbasis}
 f_{1}^{T_1} & = \eps^3 \operatorname{tr}_5 \; \mathcal{I}^{T_1,[6-2\eps]}_{1,1,1,1,1}, \nn\\
 f_{2}^{T_1} & = \eps^2 \; 2 d_{23} s_{34} \; \mathcal{I}^{T_1,[4-2\eps]}_{0,1,1,1,1}, \nn \\
 f_{3}^{T_1} & = \eps^2 \; s_{34} s_{45}   \; \mathcal{I}^{T_1,[4-2\eps]}_{1,0,1,1,1}, \nn \\
 f_{4}^{T_1} & = \eps^2 \; 2 d_{15} s_{45} \; \mathcal{I}^{T_1,[4-2\eps]}_{1,1,0,1,1}, \nn \\
 f_{5}^{T_1} & = \eps^2 \; 2 d_{15} s_{12} \; \mathcal{I}^{T_1,[4-2\eps]}_{1,1,1,0,1}, \nn \\
 f_{6}^{T_1} & = \eps^2 \; 2 d_{23} s_{12} \; \mathcal{I}^{T_1,[4-2\eps]}_{1,1,1,1,0}, \nn \\
 f_{7}^{T_1} & = \eps^2 \; \Delta_3(p_{23},p_1) \; \mathcal{I}^{T_1,[4-2\eps]}_{1,1,0,1,0}, \nn \\
 f_{8}^{T_1} & = \eps^2 \; \Delta_3(p_{15},p_2) \; \mathcal{I}^{T_1,[4-2\eps]}_{0,1,1,0,1}, \nn \\
 f_{9}^{T_1} & = \eps^2 \; \beta(s_{12},m_t^2)\;\mathcal{I}^{T_1,[4-2\eps]}_{1, 1, 1, 0, 0}, \nn \\
 f_{10}^{T_1} & = \eps \; s_{12} \; \mathcal{I}^{T_1,[4-2\eps]}_{2, 0, 1, 0, 0}, \nn \\
 f_{11}^{T_1} & = \eps \; s_{23} \; \mathcal{I}^{T_1,[4-2\eps]}_{0, 2, 0, 1, 0}, \nn \\
 f_{12}^{T_1} & = \eps \; s_{34} \; \mathcal{I}^{T_1,[4-2\eps]}_{0, 0, 2, 0, 1}, \nn \\
 f_{13}^{T_1} & = \eps \; s_{45} \; \mathcal{I}^{T_1,[4-2\eps]}_{2, 0, 0, 1, 0}, \nn \\
 f_{14}^{T_1} & = \eps \; s_{15} \; \mathcal{I}^{T_1,[4-2\eps]}_{0, 2, 0, 0, 1}, \nn \\
 f_{15}^{T_1} & = \eps \; m_t^2 \mathcal{I}^{T_1,[4-2\eps]}_{1, 2, 0, 0, 0}.
\end{align}
We compute analytically the boundary conditions for all the MIs in this topology, for which we obtain expressions in terms of MPL functions. Then we use \textsc{GiNaC} to evaluate them numerically with high precision (100 digits). The integral $f_{15}^{T_1}$ is equivalent to a tadpole, and therefore we can write its boundary value exactly:
\begin{equation}
f_{15}^{T_1} \vert_{x_0} = \frac{1}{2}e^{\eps \gamma_E} \Gamma(1+\eps)= \frac{1}{2} + \frac{\pi^2}{24}\eps^2 - \frac{\zeta_3}{6}\eps^3 + \frac{\pi^4}{320}\eps^4 + \mathcal{O}(\eps^5).
\end{equation} 

\subsection{The $T_2$ topology with 4 massive internal propagators}

\begin{figure}[h]
  \begin{center}
    \includegraphics[scale=0.45]{figs/T2all.pdf}
  \end{center}
  \caption{The 21 master integrals appearing in the $T_2$ topology, denoted by $\cI$'s in~\eqref{eq:T2-utbasis}. Red/black lines indicate massive/massless particles respectively. Dotted internal lines indicate propagators with an additional power in the denominator of the integral. Each integral is associated with a kinematic normalisation which ensures the basis leads to a canonical form differential equation.}
  \label{fig:T2-mis}
\end{figure}

Topology $T_2$ is described by 21 master integrals as shown in Figure \ref{fig:T2-mis}. The integrals are defined as:
\begin{equation}
   \label{Topo2}
   \mathcal{I}^{T_2,[d]}_{a_1,a_2,a_3,a_4,a_5} = \int \mathcal{D}^d k_1 \dfrac{1}{D_1^{a_1}D_2^{a_2}D_3^{a_3}D_4^{a_4}D_5^{a_5}},
\end{equation}
where
\begin{align}
  D_1 &= k_1^2 - m_t^2, \; D_2 = (k_1-p_1)^2, \; D_3 = (k_1 -p_1 -p_2)^2 - m_t^2, \nonumber\\
  D_4 &= (k_1 + p_4 + p_5)^2 - m_t^2, \; D_5 = (k_1 + p_5)^2 - m_t^2.
\end{align}
The canonical basis for the topology $T_2$ is chosen to be:
\begin{align}
\label{eq:T2-utbasis}
 f_{1}^{T_2} & = \eps^3 \operatorname{tr}_5 \; \mathcal{I}^{T_2,[6-2\eps]}_{1,1,1,1,1}, \nn\\
 f_{2}^{T_2} & = \eps^2 \; 4 d_{34} d_{23} \beta\left(\frac{2 d_{23} d_{34}}{d_{23} - d_{15}}, m_t^2\right) \; \mathcal{I}^{T_2,[4-2\eps]}_{0,1,1,1,1}, \nn \\
 f_{3}^{T_2} & = \eps^2 \; 4 d_{34} d_{45} \beta\left(-\frac{2 d_{45} d_{34}}{d_{35}}, m_t^2\right)   \; \mathcal{I}^{T_2,[4-2\eps]}_{1,0,1,1,1}, \nn \\
 f_{4}^{T_2} & = \eps^2 \; 4 d_{15} d_{45} \beta\left(\frac{2 d_{15} d_{45}}{d_{15} - d_{23}}, m_t^2\right) \; \mathcal{I}^{T_2,[4-2\eps]}_{1,1,0,1,1}, \nn \\
 f_{5}^{T_2} & = \eps^2 \; 2 d_{15} s_{12} \beta\left(s_{12}, m_t^2\right) \; \mathcal{I}^{T_2,[4-2\eps]}_{1,1,1,0,1}, \nn \\
 f_{6}^{T_2} & = \eps^2 \; 2 d_{23} s_{12} \beta\left(s_{12}, m_t^2\right) \; \mathcal{I}^{T_2,[4-2\eps]}_{1,1,1,1,0}, \nn \\
 f_{7}^{T_2} & = \eps^2 \; 2 d_{34} \;\mathcal{I}^{T_2,[4-2\eps]}_{0, 0, 1, 1, 1}, \nn \\
 f_{8}^{T_2} & = \eps^2 \; 2 d_{45}  \; \mathcal{I}^{T_2,[4-2\eps]}_{1,0,0,1,1}, \nn \\
 f_{9}^{T_2} & = \eps^2 \; 2 d_{23} \; \mathcal{I}^{T_2,[4-2\eps]}_{0,1,1,1,0}, \nn \\
 f_{10}^{T_2} & = \eps^2 \; 2 d_{15} \; \mathcal{I}^{T_2,[4-2\eps]}_{1, 1, 0, 0, 1}, \nn \\
 f_{11}^{T_2} & = \eps^2 \; \Delta_3\left(p_{23},p_1\right) \; \mathcal{I}^{T_2,[4-2\eps]}_{1, 1, 0, 1, 0}, \nn \\
 f_{12}^{T_2} & = \eps^2 \; \Delta_3\left(p_{15},p_2\right) \; \mathcal{I}^{T_2,[4-2\eps]}_{0, 1, 1, 0, 1}, \nn \\
 f_{13}^{T_2} & = \eps^2 \; 2(d_{12} - d_{45} + m_t^2) \; \mathcal{I}^{T_2,[4-2\eps]}_{1, 0, 1, 1, 0}, \nn \\
 f_{14}^{T_2} & = \eps^2 \; 2(d_{12} - d_{34} + m_t^2) \; \mathcal{I}^{T_2,[4-2\eps]}_{1, 0, 1, 0, 1}, \nn \\
 f_{15}^{T_2} & = \eps^2 \; 2(d_{15}  -d_{23})\mathcal{I}^{T_2,[4-2\eps]}_{0, 1, 0, 1, 1}, \nn \\
 f_{16}^{T_2} & = \eps \; s_{12} \beta\left(s_{12}, m_t^2\right) \mathcal{I}^{T_2,[4-2\eps]}_{1, 0, 2, 0, 0}, \nn \\
 f_{17}^{T_2} & = \eps \; s_{45} \beta\left(s_{45}, m_t^2\right) \mathcal{I}^{T_2,[4-2\eps]}_{2, 0, 0, 1, 0}, \nn \\
 f_{18}^{T_2} & = \eps \; s_{34} \beta\left(s_{34}, m_t^2\right) \mathcal{I}^{T_2,[4-2\eps]}_{0, 0, 1, 0, 2}, \nn \\
 f_{19}^{T_2} & = \eps \; s_{23} \mathcal{I}^{T_2,[4-2\eps]}_{0, 1, 0, 2, 0}, \nn \\
 f_{20}^{T_2} & = \eps \; s_{15} \mathcal{I}^{T_2,[4-2\eps]}_{0, 1, 0, 0, 2}, \nn \\
 f_{21}^{T_2} & = \eps \; m_t^2 \mathcal{I}^{T_2,[4-2\eps]}_{2, 1, 0, 0, 0}.
\end{align}
As for topology $T_1$, we compute analytically the boundary conditions for all
the MIs in topology $T_2$ but for the pentagon $f_1^{T_2}$, for which we obtain
an expression in terms of one-parameter integrals. Moreover, since
\begin{equation}
f_{21}^{T_2} = f_{15}^{T_1}, \,\,\, f_{20}^{T_2} = f_{14}^{T_1}, \,\,\, f_{19}^{T_2} = f_{11}^{T_1},
\end{equation}
we do not have to perform any new computation for these integrals.

\subsection{The $T_3$ topology with 2 massive internal propagators}

\begin{figure}[h]
  \begin{center}
    \includegraphics[scale=0.45]{figs/T3all.pdf}
  \end{center}
  \caption{The 17 master integrals appearing in topology $T_3$, denoted by $\cI$'s in~\eqref{eq:T3-utbasis}. Red/black lines indicate massive/massless particles respectively. Dotted internal lines indicate propagators with an additional power in the denominator of the integral. Each integral is associated with a kinematic normalisation which ensures the basis leads to a canonical form differential equation.}
  \label{fig:T3-mis}
\end{figure}

Topology $T_3$ is described by 17 master integrals as shown in Figure \ref{fig:T3-mis}. The integrals are defined as:
\begin{equation}
   \label{Topo3}
   \mathcal{I}^{T_3,[d]}_{a_1,a_2,a_3,a_4,a_5} = \int \mathcal{D}^d k_1 \dfrac{1}{D_1^{a_1}D_2^{a_2}D_3^{a_3}D_4^{a_4}D_5^{a_5}},
\end{equation}
where
\begin{align}
  D_1 &= k_1^2, \; D_2 = (k_1-p_1)^2 - m_t^2, \; D_3 = (k_1 -p_1 -p_3)^2 - m_t^2, \nonumber\\
  D_4 &= (k_1 + p_4 + p_5)^2, \; D_5 = (k_1 + p_5)^2.
\end{align}
The canonical basis for the topology $T_3$ is chosen to be:
\begin{align}
\label{eq:T3-utbasis}
 f_{1}^{T_3} & = \eps^3 \operatorname{tr}_5 \; \mathcal{I}^{T_3,[6-2\eps]}_{1,1,1,1,1}, \nn\\
 f_{2}^{T_3} & = \eps^2 \; 4 d_{24} d_{23} \; \mathcal{I}^{T_3,[4-2\eps]}_{0,1,1,1,1}, \nn \\
 f_{3}^{T_3} & = \eps^2 \; 4 d_{24} d_{45} \; \mathcal{I}^{T_3,[4-2\eps]}_{1,0,1,1,1}, \nn \\
 f_{4}^{T_3} & = \eps^2 \; 4 d_{15} d_{45} \; \mathcal{I}^{T_3,[4-2\eps]}_{1,1,0,1,1}, \nn \\
 f_{5}^{T_3} & = \eps^2 \; 4 d_{15} d_{13} \; \mathcal{I}^{T_3,[4-2\eps]}_{1,1,1,0,1}, \nn \\
 f_{6}^{T_3} & = \eps^2 \; 4 d_{23} d_{13} \beta\left(-\frac{2 d_{13} d_{23}}{d_{45}}, m_t^2\right) \; \mathcal{I}^{T_3,[4-2\eps]}_{1,1,1,1,0}, \nn \\
 f_{7}^{T_3} & = \eps^2 \; \Delta_3\left(p_{13}, p_2\right) \;\mathcal{I}^{T_3,[4-2\eps]}_{1, 0, 1, 1, 0}, \nn \\
 f_{8}^{T_3} & = \eps^2 \; \Delta_3\left(p_{15}, p_2\right) \; \mathcal{I}^{T_3,[4-2\eps]}_{1,1,0,1,0}, \nn \\
 f_{9}^{T_3} & = \eps^2 \; 2 d_{23} \; \mathcal{I}^{T_3,[4-2\eps]}_{0,1,1,1,0}, \nn \\
 f_{10}^{T_3} & = \eps^2 \; 2 d_{13} \; \mathcal{I}^{T_3,[4-2\eps]}_{1, 1, 1, 0, 0}, \nn \\
 f_{11}^{T_3} & = \eps^2 \; 2 (d_{15} - d_{24})\; \mathcal{I}^{T_3,[4-2\eps]}_{0, 1, 1, 0, 1}, \nn \\
 f_{12}^{T_3} & = \eps \; s_{13} \; \mathcal{I}^{T_3,[4-2\eps]}_{1, 0, 2, 0, 0}, \nn \\
 f_{13}^{T_3} & = \eps \; s_{23} \; \mathcal{I}^{T_3,[4-2\eps]}_{0, 2, 0, 1, 0}, \nn \\
 f_{14}^{T_3} & = \eps \; s_{24} \; \mathcal{I}^{T_3,[4-2\eps]}_{0, 0, 2, 0, 1}, \nn \\
 f_{15}^{T_3} & = \eps \; s_{45} \mathcal{I}^{T_3,[4-2\eps]}_{2, 0, 0, 1, 0}, \nn \\
 f_{16}^{T_3} & = \eps \; s_{15} \mathcal{I}^{T_3,[4-2\eps]}_{0, 2, 0, 0, 1}, \nn \\
 f_{17}^{T_3} & = \eps \; m_t^2 \mathcal{I}^{T_2,[4-2\eps]}_{1, 2, 0, 0, 0}.
\end{align}
The boundary conditions for the bubble integrals $f_{17}^{T_3},\,
f_{16}^{T_3},\, f_{15}^{T_3},\, f_{13}^{T_3}$ are known from topology $T_1$,
while the values for the integrals $f_{14}^{T_3},\, f_{12}^{T_3}$ can be
obtained numerically using \textsc{DiffExp}. For example, the boundary
condition for $f_{14}^{T_3}$ can be obtained from $f_{13}^{T_3}$ by evolving,
using \textsc{DiffExp}, its value at $\vec{x}_0$ to a point
$\vec{x}_0^{\sigma}$, whose value is determined by an appropriate permutation
of the external momenta. Indeed, $f_{14}^{T_3}$ is the same integral as
$f_{13}^{T_3}$ just in a different channel, as it can be seen from Figure
\ref{fig:T3-mis}. We also exploit this strategy for other integrals in $T_3$ and $T_4$.

In $T_3$, only the boundary values for the integrals $f_{1}^{T_3},\,
f_{2}^{T_3},\, f_{6}^{T_3}$ and $f_{9}^{T_3}$ need to be computed explicitly
through direct integration. All other boundary values have already been
computed in previous topologies or through numerical evaluation using
\textsc{DiffExp}. In particular, the integrals $f_{8}^{T_3}$ and $f_{4}^{T_3}$
appear already in other topologies,
\begin{equation}
f_{8}^{T_3} = f_{7}^{T_1}, \,\,\, f_{4}^{T_3} = f_{4}^{T_1} \,,
\end{equation}
and the boundary values for the remaining MIs are evaluated with \textsc{DiffExp} in the following way:
\begin{itemize}
\item the boundary value for $f_{11}^{T_3}$ is obtained from $f_{15}^{T_2}$;
\item the boundary value for $f_{10}^{T_3}$ is obtained from $f_{9}^{T_2}$;
\item the boundary value for $f_{7}^{T_3}$ is obtained from $f_{8}^{T_1}$;
\item the boundary value for $f_{5}^{T_3}$ is obtained from $f_{2}^{T_3}$;
\item the boundary value for $f_{3}^{T_3}$ is obtained from $f_{4}^{T_1}$.
\end{itemize}  

\subsection{The $T_4$ topology with 3 massive internal propagators}

\begin{figure}[h]
  \begin{center}
    \includegraphics[scale=0.45]{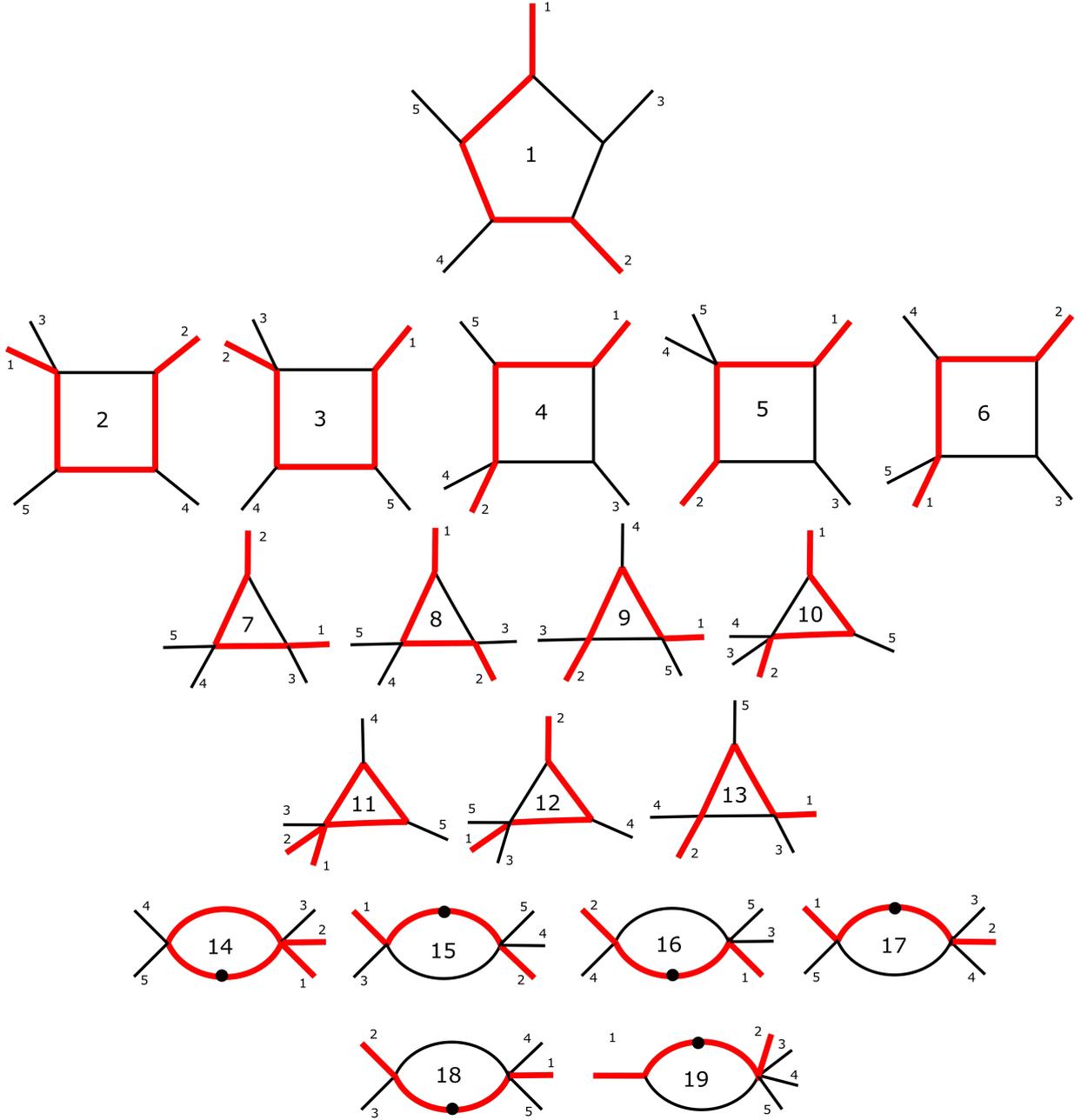}
  \end{center}
  \caption{The 19 master integrals appearing in topology $T_4$, denoted by $\cI$'s in~\eqref{eq:T4-utbasis}. Red/black lines indicate massive/massless particles respectively. Dotted internal lines indicate propagators with an additional power in the denominator of the integral. Each integral is associated with a kinematic normalisation which ensures the basis leads to a canonical form differential equation.}
  \label{fig:T4-mis}
\end{figure}

Topology $T_4$ is described by 19 master integrals as shown in Figure \ref{fig:T4-mis}. The integrals are defined as:
\begin{equation}
   \label{Topo4}
   \mathcal{I}^{T_3,[d]}_{a_1,a_2,a_3,a_4,a_5} = \int \mathcal{D}^d k_1 \dfrac{1}{D_1^{a_1}D_2^{a_2}D_3^{a_3}D_4^{a_4}D_5^{a_5}},
\end{equation}
where
\begin{align}
  D_1 &= k_1^2 - m_t^2, \; D_2 = (k_1-p_1)^2, \; D_3 = (k_1 -p_1 -p_3)^2, \nonumber\\
  D_4 &= (k_1 + p_4 + p_5)^2 - m_t^2, \; D_5 = (k_1 + p_5)^2 - m_t^2.
\end{align}

The canonical basis for the topology $T_4$ is chosen to be:
\begin{align}
\label{eq:T4-utbasis}
 f_{1}^{T_4} & = \eps^3 \operatorname{tr}_5 \; \mathcal{I}^{T_4,[6-2\eps]}_{1,1,1,1,1}, \nn\\
 f_{2}^{T_4} & = \eps^2 \; 4 d_{24} d_{45} \beta\left(\frac{2 d_{24} d_{45}}{d_{24} - d_{13}}, m_t^2\right) \; \mathcal{I}^{T_4,[4-2\eps]}_{1,0,1,1,1}, \nn \\
 f_{3}^{T_4} & = \eps^2 \; 4 d_{15} d_{45} \beta\left(\frac{2 d_{23} d_{45}}{d_{23} - d_{15}}, m_t^2\right) \; \mathcal{I}^{T_4,[4-2\eps]}_{1,1,0,1,1}, \nn \\
 f_{4}^{T_4} & = \eps^2 \; 4 d_{15} d_{13} \; \mathcal{I}^{T_4,[4-2\eps]}_{1,1,1,0,1}, \nn \\
 f_{5}^{T_4} & = \eps^2 \; 4 d_{23} d_{13} \; \mathcal{I}^{T_4,[4-2\eps]}_{1,1,1,1,0}, \nn \\
 f_{6}^{T_4} & = \eps^2 \; 4 d_{24} d_{23} \; \mathcal{I}^{T_4,[4-2\eps]}_{0,1,1,1,1}, \nn \\
 f_{7}^{T_4} & = \eps^2 \; \Delta_3\left(p_{13}, p_2\right) \;\mathcal{I}^{T_4,[4-2\eps]}_{1, 0, 1, 1, 0}, \nn \\
 f_{8}^{T_4} & = \eps^2 \; \Delta_3\left(d_{23}, p_1\right) \; \mathcal{I}^{T_4,[4-2\eps]}_{1,1,0,1,0}, \nn \\
 f_{9}^{T_4} & = \eps^2 \; 2 (d_{23}-d_{15}) \; \mathcal{I}^{T_4,[4-2\eps]}_{0,1,0,1,1}, \nn \\
 f_{10}^{T_4} & = \eps^2 \; 2 d_{15} \; \mathcal{I}^{T_4,[4-2\eps]}_{1, 1, 0, 0, 1}, \nn \\
 f_{11}^{T_4} & = \eps^2 \; 2 d_{45}\; \mathcal{I}^{T_4,[4-2\eps]}_{1, 0, 0, 1, 1}, \nn \\
 f_{12}^{T_4} & = \eps^2 \; 2 d_{24} \; \mathcal{I}^{T_4,[4-2\eps]}_{0, 0, 1, 1, 1}, \nn \\
 f_{13}^{T_4} & = \eps^2 \; 2(d_{13} - d_{24}) \; \mathcal{I}^{T_4,[4-2\eps]}_{1, 0, 1, 0, 1}, \nn \\
 f_{14}^{T_4} & = \eps \; s_{45} \beta\left( s_{45}, m_t^2 \right) \; \mathcal{I}^{T_4,[4-2\eps]}_{1, 0, 0, 2, 0}, \nn \\
 f_{15}^{T_4} & = \eps \; s_{13} \mathcal{I}^{T_4,[4-2\eps]}_{2, 0, 1, 0, 0}, \nn \\
 f_{16}^{T_4} & = \eps \; s_{24} \mathcal{I}^{T_4,[4-2\eps]}_{0, 0, 1, 0, 2}, \nn \\
 f_{17}^{T_4} & = \eps \; s_{15} \mathcal{I}^{T_4,[4-2\eps]}_{0, 1, 0, 0, 2}, \nn \\
 f_{18}^{T_4} & = \eps \; s_{23} \mathcal{I}^{T_4,[4-2\eps]}_{0, 1, 0, 2, 0}, \nn \\
 f_{19}^{T_4} & = \eps \; m_t^2 \mathcal{I}^{T_2,[4-2\eps]}_{2, 1, 0, 0, 0}.
\end{align}
The boundary conditions for all the bubble and the triangle integrals, and most
of the box integrals too, have already been considered in the previous
topologies. The only new MIs for which we compute the boundary values by direct
integration are $f_{1}^{T_4}$ and $f_{5}^{T_4}$.

\section{Results \label{sec:res}}

\subsection{Numerical results for the master integrals \label{sec:res_mis}}

In this section we discuss our results for the numerical evaluation of the MIs
performed with \textsc{DiffExp}. The amplitudes depend on 130 independent MIs
across all the four pentagon topologies, and their permutations. Instead of
evaluating the whole system of 130 MIs at once we evaluate each topology
separately. Since the number of MIs inside each topology is at most 21, this
approach allows us a faster numerical evaluation, as we can evaluate in
parallel all the topologies and their permutations. The timing to get numerical values
for all the topologies and permutations is within a range of $\sim 30$ minutes to $\sim 1$ hour for phase-space point, on a laptop, requiring an accuracy of 16 digits. We stress that for phenomenological applications the performances can be improved by building a precomputed grid of points as boundary values \cite{Abreu:2020jxa,Becchetti:2020wof}.\par Given a point
$\vec{x}_a$, we can evaluate all the MIs, and hence the amplitude, at that
point as follows. The standard ordering of the external momenta for topologies
$T_1$ and $T_2$ is $(1,2,3,4,5)$, while for topologies $T_3$ and $T_4$ it is
$(1,3,2,4,5)$. The permutations of these topologies are given by all the
possible permutations of the momenta $(3,4,5)$. Therefore, evaluating the MIs
that belong to the permutations of $T_1$,$T_2$,$T_3$ and $T_4$ is equivalent to
evaluating the MIs that describe the topologies in the standard orderings
$(1,2,3,4,5)$ and $(1,3,2,4,5)$ at a kinematic point, $\vec{x}_a^{\sigma}$,
which is given by the corresponding permutations of the kinematic invariants.
\par In order to clarify this procedure we discuss the following example. We
consider the permutation $(1,2,4,3,5)$ for the topology $T_1$. The permutation
of the external momenta:
\begin{equation}
p_3 \rightarrow p_4, \,\,\, p_4 \rightarrow p_3
\end{equation}
implies the following transformation for the kinematics invariants:
\begin{align}
\bigg\{
  &d_{12}\to d_{12},d_{23}\to d_{15}-d_{23}-d_{34},\nonumber\\&
  d_{34}\to d_{34},d_{45}\to d_{12}-d_{34}-d_{45}+m_t^2,d_{15}\to d_{15}
\bigg\}.
\end{align}
This means that evaluating the permutation $(1,2,4,3,5)$ of $T_1$ at the point:
\begin{equation}
\vec{x}_a = \left\{d_{12} \to -\tfrac{11}{7}, d_{23} \to -\tfrac{7}{5}, d_{34} \to -\tfrac{5}{27}, d_{45} \to -\tfrac{17}{5}, d_{15} \to -\tfrac{11}{17}, m_t^2 \to 1\right\}
\end{equation}
is equivalent to evaluating $T_1$, in the standard ordering $(1,2,3,4,5)$, at the point:
\begin{equation}
\vec{x}_a^{\sigma} = \left\{d_{12} \to -\tfrac{11}{7}, d_{23} \to \tfrac{2153}{2295}, d_{34} \to -\tfrac{5}{27}, d_{45} \to \tfrac{2848}{945}, d_{15} \to -\tfrac{11}{17}, m_t^2 \to 1\right\}.
\end{equation}
This procedure allows us to evaluate all the permutations of a given topology
starting from the system of differential equations for the topology in the
standard ordering. Moreover, this strategy has also been used to compute the
boundary conditions for some MIs as discussed in the previous section.\par In
order to verify the correctness of our computation we performed different
checks comparing our results against numerical values for the MIs obtained by
means of sector decomposition techniques, as implemented in the software
\textsc{PySecDec} \cite{Borowka:2017idc}.

\subsection{Amplitude results \label{sec:res_amp}}

The explicit analytic forms for the partial helicity amplitudes, broken into
subamplitudes according to Eq. \eqref{eq:spindecomp}, are provided in the ancillary
files. Due to the large overall size we do not attempt to provide any typeset
expressions in the paper. The coefficients appearing in the subamplitudes have
been collected and linear relations between them determined. To provide a
relatively compact format common factors in the linearly indpendent rational
coefficients are identified and presented as a set of replacement rules.

The univariate partial fractioning in the variable $x_{5123}$ was quite
effective in reducing the total number of sample points required in the
reconstruction. The maximum total degree appearing in the most complicated sub-leading
colour amplitudes was $\mathcal{O}(100)$ before partial fractioning and linear
relations. This reduced to $\mathcal{O}(20)$ in the final expressions.
Nevertheless, the appearance of high degree polynomials does indicate that
extensions to higher loops will be challenging since the fast evaluation of the
one-loop input enables to handle such expressions without restrictions on
computational resources.

The ancilary files also provide two example scripts demonstrating the
evaluation of the amplitudes using the numerical results for the master
integrals and validation of the universal pole structure. All evaluations have
been performed in \textsc{Mathematica} where we can obtain~$\mathcal{O}(100)$
accurate digits without issues. Since the main use case of the new
$\mathcal{O}(\eps)$ and $\mathcal{O}(\eps^2)$ terms will be in the subtraction
of divergences in two-loop amplitudes, we have not attempted to provide an efficient
evaluation of the amplitudes for use at NLO.

\section{Conclusions}

In this article we have presented a computation of all one-loop helicity
amplitudes of the process $pp\to t\bar{t} j$ evaluated to
$\mathcal{O}(\eps^2)$.  The expansion to higher order in $\eps$ allows us to
look at the complexity of the NNLO terms for the first time. Applying finite
field reconstruction techniques demonstrates that the algebraic complexity of
this problem may be within reach. The analytic complexity coming from the loop
integrals was easily overcome using the combination of canonical form
differential equations and subsequent evaluation using generalised series
expansions in \textsc{DiffExp}. The boundary terms are provided in analytic
form up to weight four using MPLs except for the pentagon master integral with four internal
masses which is presented as a one-paramater integral. 

It will be interesting to see how automated approaches to loop integral
evaluation using the numerical evaluation of the differential equations
develop. Since mathematical bottlenecks in the understanding of elliptic
structures and the difficulties of dealing with long and complicated alphabets
may be sidestepped, the method has substantial advantages over fully analytic
approaches. Nevertheless this comes at the cost of numerical performance and
the determination of the boundary values will still be a major issue. Recent
attempts to automate the evaluation of boundary terms using sector
decomposition have been successful~\cite{Dubovyk:2022frj} although the
numerical accuracy is probably not yet sufficient for the full phase space.

There are clearly important issues that should be addressed in order to
overcome challenges at two-loops. Amplitudes in $d=4-2\eps$ dimensions are
substantially more complicated than their four-dimensional limits.  The
identification of an analytic function basis such that expansion in $\eps$ can
be taken and the subtraction of poles can be performed is likely to be an
essential ingredient. We also observe a high degree of algebraic complexity
stemming from the global choice of rational kinematic parametrisation. This is
particularly evident in the sub-leading colour partial amplitudes in which many
permutations of the master integral topologies appear.

Despite significant challenges ahead, the work presented here motivates further
investigation into analytic or semi-analytic approaches to high precision
$pp\to t\tb j$ amplitudes and cross-sections.

\section{Acknowledgements}

We are grateful to Bayu Hartanto for many useful discussions and comments on
the manuscript. We are also grateful to Simone Zoia, Simone Alioli and Armin Schweitzer for
helpful discussions. This project received funding from the European Union's
Horizon 2020 research and innovation programmes \textit{High precision
multi-jet dynamics at the LHC} (consolidator grant agreement No 772009) and
\textit{Manifesting the Simplicity of Scattering Amplitudes} (starting grant
agreement No 757978). RM acknowledges additional support from the Villum Fonden
research grant 00025445.
\appendix

\section{Generalized series expansion method: A brief review}

For the reader's convenience, we start this section with a short review of the method of generalized power series \cite{Francesco:2019yqt, Abreu:2020jxa}, implemented in \cite{Hidding:2020ytt}, which has been exploited to evaluate the MIs numerically. \par 
The generalized series expansion method allows us to evaluate the solution to the system \eqref{eq:deqsCan}, at a point $\vec{x}_a$, from the knowledge of the solution at some boundary point $\vec{x}_0$. This is done is the following three steps:
\begin{itemize}
\item \textbf{Step 1}: We split the integration path into segments;
\item \textbf{Step 2}: We find a solution inside each segment by expanding in series the system of differential equations;
\item \textbf{Step 3}: We evaluate the solution in the point $\vec{x}_a$ connecting the local solutions of the path $\gamma(t)$.
\end{itemize}
The solution to the canonical system \eqref{eq:deqsCan} can be written as a series expansion in $\eps$:
\begin{equation} 
\vec{f}(t, \eps) = \sum_{k = 0}^{\infty} \eps^k \vec{f}^{(k)} (t),
\end{equation}
where:
\begin{equation} \label{eq:wksol}
\vec{f}^{(k)}(t) = \sum_{j = 1}^{k}\int_0^{1}d t_1 A(t_1)\int_0^{t_1} d t_2 A(t_2)\cdots\int_0^{t_{j-1}}d t_j A(t_j)\vec{f}^{(k-j)}(\vec{x}_0)+\vec{f}^{(k)}(\vec{x}_0),  
\end{equation}
and we assume that the solution is described by some variable $t$ which parametrizes the path, $\gamma(t)$, which connects the points $\vec{x}_0$ and $\vec{x}_a$:
\begin{equation}
\gamma(t) \, : \, t \mapsto \vec{x} (t), \,\,\, t \in \left[0,1\right], \,\,\, \gamma(0) = \vec{x}_0 \, , \, \gamma(1) = \vec{x}_a.
\end{equation}
As already mentioned, the first step consists in splitting the path $\gamma(t)$ into segments $S_i \equiv \left[t_i - r_i, t_i + r_i\right)$, where $\left\{t_i\right\}$ is the set of points in which we are going to expand the system of differential equations, and $r_i$ is the radius of convergence of the series inside each segment. The segments $S_i$ can be chosen from the knowledge of the singular points of the differential equations. In particular we can have both real:
\begin{equation}
R \equiv \left\{ \tau_i  \,\,\, \vert \,\,\, i = 1,\cdots,N_r \right\},
\end{equation}
and complex-valued singular points:
\begin{equation}
C \equiv \left\{\lambda_i^{re} + i \lambda_i^{im} \,\,\, \vert \,\,\, i = 1,\cdots, N_c\right\}.
\end{equation} 
Therefore, we can choose the expansion points to belong to the set $R \cup C_r$, where $C_r$ is a set of regular points:
\begin{equation}
C_r \equiv \cup_{i=1}^{N_c} \left\{\lambda_i^{re} \pm \lambda_i^{im} \right\}
\end{equation} 
 and the radius of convergence, $r_i$, can be defined as the distance of $t_i$ to the closest element $t_p$, with $p \neq i$.\par
In the second step of the method we determine local solutions to the differential equations inside each segment $S_i$. This is done by expanding the system of differential equations around the point $t_i$:
\begin{equation} \label{eq:deqsexp}
A(t) = \sum_{l=0}^{\infty} A_l \left(t-t_i\right)^{w_l}, \,\,\, w_l \in \mathbb{Q},    
\end{equation}
where $A_l$ are constant matrices. Then, exploiting the general expression \eqref{eq:wksol} for the $k$-weight of the solution, we obtain:
\begin{align} \label{eq:local1}
\vec{f}_i^{(k)}(t) =& \sum_{j=1}^k \sum_{l_1=0}^{\infty}\cdots\sum_{l_j=0}^{\infty} A_{l_1} \cdots A_{l_j} \int_0^t d t_1\left(t_1-t_i\right)^{w_{l_1}}\cdots \int_0^{t_{j-1}} d t_j \left(t_j - t_i\right)^{w_{l_j}}\vec{f}_i^{(k-j)}(\vec{x}_0) \nonumber \\
&+\vec{f}_i^{(k)}(\vec{x}_0).
\end{align}
for the local solution, $\vec{f}^{(k)}_i (t)$, inside the segment $S_i$.\par
We point out that working with a system of differential equations in canonical form implies that the integrals that appear in \eqref{eq:local1} are of the form:
\begin{equation}
\int_0^{t_0} d t \left(t - t_i\right)^w \log\left(t - t_i\right)^m, \,\,\, w\in \mathbb{Q}, \,\, m\in \mathbb{N}.    
\end{equation}
Consequently, \eqref{eq:local1} is given by the expression:
\begin{equation}
\vec{f}_i^{(k)}(t) = \sum_{l_1 = 0}^{\infty}\sum_{l_2 = 0}^{N_{i,k}} c_k^{(i,l_1,l_2)} \left(t - t_i\right)^{\frac{l_1}{2}} \log(t-t_i)^{l_2}, \label{eq:expanded_laurent_coefficient}
\end{equation}
where the matrices $c_k^{(i,l_1,l_2)}$ depend on the boundary conditions for the system and on the constant matrices $A_l$ in \eqref{eq:deqsexp}.\par
Finally, as last step of the procedure, the global solution on the path $\gamma(t)$ can then be approximated as:
\begin{align} \label{eq:solutionexpanded}
\vec{f}(t, \eps) = \sum_{k=0}^{\infty} \eps^k \sum_{i = 0}^{N - 1}\rho_i (t) \vec{f}^{(k)}_i (t), \;\;\; \rho(t)  = 
\begin{cases}
1, & t \in \left[t_i -r_i, t_i + r_i\right) \\
0, & t \notin \left[t_i -r_i, t_i + r_i\right)
\end{cases},
\end{align}
where $N$ is the total number of segments, and $\vec{f}^{(k)}_i (t)$ is the $k$-weight of the local solution, inside the segment $S_i$, written as a truncated series expansion, around some point $t_i$, with radius of convergence $r_i$.

\subsection{Analytic continuation}

In this part we briefly discuss the analytic continuation within the framework of the generalized series expansion method and, in particular, in its \textsc{DiffExp} implementation. \par
Analytic continuation has to be performed when a singularity of the differential equations matrix, $dA(\vec{x})$, is crossed along the integration from the boundary point, $\vec{x}_0$, to the evaluation point $\vec{x}_a$. We encounter two kinds of singularities:
\begin{itemize}
\item \textbf{Case I}: \emph{Logarithmic singularities}, this type of singularities arise from simple poles in the system of differential equations;
\item \textbf{Case II}: \emph{Square roots singularities}, this type of singularities appear as the canonical basis of MIs involves square roots of the kinematic invariants.
\end{itemize}
In order to determine all possible logarithmic singularities, we consider the system of differential equations for a basis of MIs without square roots normalization involved, then we perform a multivariate partial fraction on the system exploiting the Mathematica package \textsc{MultivariateApart} \cite{Heller:2021qkz}. In doing so, we obtain a set of irreducible polynomials in the kinematic invariants, $\mathcal{P}(\vec{x})$, which describes the simple poles structure of the system \eqref{eq:deqsexp}. Instead, the square roots singularities, $\mathcal{S}(\vec{x})$, are given by set of square roots \eqref{eq:sqrt} that define the canonical basis of MIs. Therefore, the full set of singularities for the analytic continuation is given by the set of polynomials $\mathcal{P}(\vec{x}) \cup \mathcal{S}(\vec{x})$.\par
Once the full set of singularities is known we perform the analytic continuation as follows. We assign a small imaginary part to the kinematic invariants and the top mass:
\begin{equation} \label{eq:dijac}
d_{ij} \rightarrow d_{ij} \pm i \delta, \,\,\, m_t^2 \rightarrow m_t^2 \pm i \delta,
\end{equation}
then we substitute \eqref{eq:dijac} into the polynomials $\mathcal{P}(\vec{x}) \cup \mathcal{S}(\vec{x})$, we expand with respect to $\delta$ and we keep just the linear term. As an example of this procedure we consider the logarithmic singularity
\begin{equation} \label{eq:exac1}
d_{12} - d_{34} - d_{45} + m_t^2
\end{equation}  
which appears in Topology 1. The kinematic invariants and top mass carry the imaginary parts
\begin{equation} \label{eq:exac2}
\left\{d_{12} + i \delta, d_{23} + i \delta, d_{34} + i \delta, d_{45} + i \delta, d_{15} + i \delta, m_t^2 - i \delta\right\}\,.
\end{equation}
As a consequence, once we substitute \eqref{eq:exac2} into \eqref{eq:exac1}, we obtain that the singularity \eqref{eq:exac1} is analytic continued as:
\begin{equation}
d_{12} - d_{34} - d_{45} + m_t^2 - i \delta .
\end{equation}

\section{Explicit form of the infrared poles for the partial amplitudes \label{app:irpoles}}

In this appendix we give the explict form for the Catani-Dittmaier-Trocsanyi
formula Eq. \eqref{eq:CDTpoles} for the partial colour amplitudes. We use the
following short hand for the logarithms that appear,
\begin{align}
  L_{ij} &= \log\left( \frac{\mu_R^2}{-2 d_{ij}} \right) \\
  L_{m,ij} &= \frac{1}{2}\log\left( \frac{\mu_R^2}{-2 d_{ij}} \right) + \frac{1}{2}\log\left( \frac{m_t^2}{-2 d_{ij}} \right)\\
  L_{\beta} &= \frac{2 d_{12}}{s_{12}\beta} \log \left( -\frac{1-\beta}{1+\beta} \right)
  \label{eq:logshorthand}
\end{align}
The poles are list to order $\eps$:
\begin{equation}
  A_x^{L,d_s} = P_x^{(L,d_s)} + \mathcal{O}(\eps^0).
\end{equation}
All forumlae are also available in comupter readable forms in the ancillary files. Firstly for $0\to t\tb ggg$ process,

\begin{align}
  P_{1;1}^{(1,0)}(1,2,3,4,5) &=  A_{1;0}^{(0)}(1,2,3,4,5)\left( -\tfrac{3}{\eps^2} - \tfrac{1}{\eps}\left( L_{m,23}+L_{m,15}+L_{34}+L_{45} \right) \right)
  \label{eq:pole_ttggg_NcT23451_ds0} \\
  P_{1;1}^{(1,1)}(1,2,3,4,5) &=  A_{1;0}^{(0)}(1,2,3,4,5)\left( \tfrac{1}{4\eps} \right)
  \label{eq:pole_ttggg_NcT23451_ds1}
\end{align}

\begin{align}
  P_{1;-1}^{(1,0)}(1,2,3,4,5) &=  A_{1;0}^{(0)}(1,2,3,4,5)\left( \tfrac{1}{\eps} L_{\beta} \right)
  \label{eq:pole_ttggg_Ncpm1T23451_ds0} \\
  P_{1;-1}^{(1,1)}(1,2,3,4,5) &=  A_{1;0}^{(0)}(1,2,3,4,5)\left( -\tfrac{1}{4\eps} \right)
  \label{eq:pole_ttggg_Ncpm1T23451_ds1}
\end{align}

\begin{align}
  P_{2;0}^{(1,0)}(1,2,3,4,5) &= 
     -A_{1;0}^{(0)}(1,2,3,5,4)\left( \tfrac{1}{\eps}\left( L_{m,13}-L_{m,15}-L_{34}+L_{45} \right)  \right) \nonumber\\&
     -A_{1;0}^{(0)}(1,2,4,5,3)\left( \tfrac{1}{\eps}\left( L_{m,23}-L_{m,25}-L_{34}+L_{45} \right)  \right) \nonumber\\&
     -A_{1;0}^{(0)}(1,2,3,4,5)\left( \tfrac{1}{\eps}\left( L_{m,13}-L_{m,14}-L_{35}+L_{45} \right)  \right) \nonumber\\&
     -A_{1;0}^{(0)}(1,2,5,4,3)\left( \tfrac{1}{\eps}\left( L_{m,23}-L_{m,24}-L_{35}+L_{45} \right)  \right)
  \label{eq:pole_ttggg_d45T231_ds0} \\
  P_{2;0}^{(1,1)}(1,2,3,4,5) &= 0
  \label{eq:pole_ttggg_d45T231_ds1}
\end{align}

\begin{align}
  P_{3;0}^{(1,0)}(1,2,3,4,5) &= 
      A_{1;0}^{(0)}(1,2,4,5,3)\left( \tfrac{1}{\eps}\left( L_{m,14}+L_{m,23}-L_{34}-L_{\beta} \right)  \right) \nonumber\\&
     +A_{1;0}^{(0)}(1,2,3,4,5)\left( \tfrac{1}{\eps}\left( L_{m,13}+L_{m,25}-L_{35}-L_{\beta} \right)  \right) \nonumber\\&
     +A_{1;0}^{(0)}(1,2,5,3,4)\left( \tfrac{1}{\eps}\left( L_{m,15}+L_{m,24}-L_{45}-L_{\beta} \right)  \right)
  \label{eq:pole_ttggg_d21TR345_ds0} \\
  P_{3;0}^{(1,1)}(1,2,3,4,5) &= 0
  \label{eq:pole_ttggg_d21TR345_ds1}
\end{align}

Closed fermion loops are finite in the $t\tb ggg$ channel. For $0\to t\tb q \qb g$ process we have:

\begin{align}
  P_{1;1}^{(1,0)}(1,2,3,4,5) &=  A_{1;0}^{(0)}(1,2,3,4,5)\left( -\tfrac{2}{\eps^2} - \tfrac{1}{\eps}\left( L_{m,14}+L_{m,25}+L_{35} - 2 \right) \right)
  \label{eq:pole_ttqqg_Ncd14T253_ds0} \\
  P_{1;1}^{(1,1)}(1,2,3,4,5) &=  A_{1;0}^{(0)}(1,2,3,4,5)\left( \tfrac{1}{3\eps} \right)
  \label{eq:pole_ttqqg_Ncd14T253_ds1}
\end{align}

\begin{align}
  P_{1;-1}^{(1,0)}(1,2,3,4,5) &=
  A_{1;0}^{(0)}(1,2,3,4,5)\big( \tfrac{1}{\eps^2} \nonumber\\& \hspace{0.5cm} + \tfrac{1}{\eps}\left( L_{m,14}+L_{m,23}-L_{m,13}-L_{m,24}+L_{34}+L_{\beta} + 2 \right) \big) \nonumber\\&
  + A_{3;0}^{(0)}(1,2,3,4,5)\left( \tfrac{1}{\eps}\left( L_{m,15}-L_{m,14}+L_{m,24}-L_{m,25} \right) \right) \nonumber\\&
  + A_{4;0}^{(0)}(1,2,3,4,5)\left( \tfrac{1}{\eps}\left( L_{45}-L_{35}+L_{m,13}-L_{m,14} \right) \right)
  \label{eq:pole_ttqqg_Ncpm1d14T253_ds0} \\
  P_{1;-1}^{(1,1)}(1,2,3,4,5) &=  A_{1;0}^{(0)}(1,2,3,4,5)\left( -\tfrac{1}{2\eps} \right)
  \label{eq:pole_ttqqg_Ncpm1d14T253_ds1}
\end{align}

\begin{align}
  P_{2;1}^{(1,0)}(1,2,3,4,5) &=
  A_{2;0}^{(0)}(1,2,3,4,5)\left( -\tfrac{2}{\eps^2} - \tfrac{1}{\eps}\left( L_{m,15}+L_{m,23}+L_{45} - 2 \right) \right)
  \label{eq:pole_ttqqg_Ncd23T451_ds0} \\
  P_{2;1}^{(1,1)}(1,2,3,4,5) &=
  A_{2;0}^{(0)}(1,2,3,4,5)\left( \tfrac{1}{3\eps} \right)
  \label{eq:pole_ttqg_Ncd23T451_ds1}
\end{align}

\begin{align}
  P_{2;-1}^{(1,0)}(1,2,3,4,5) &=
    A_{2;0}^{(0)}(1,2,3,4,5)\big( \tfrac{1}{\eps^2} \nonumber\\& \hspace{0.5cm} + \tfrac{1}{\eps}\left( L_{m,14}+L_{m,23}-L_{m,13}-L_{m,24}+L_{34}+L_{\beta} + 2 \right) \big) \nonumber\\&
  + A_{3;0}^{(0)}(1,2,3,4,5)\left( \tfrac{1}{\eps}\left( L_{m,13}-L_{m,15}+L_{m,25}-L_{m,23} \right) \right) \nonumber\\&
  + A_{4;0}^{(0)}(1,2,3,4,5)\left( \tfrac{1}{\eps}\left( L_{35}-L_{45}+L_{m,24}-L_{m,23} \right) \right)
  \label{eq:pole_ttqqg_Ncpm1d23T451_ds0} \\
  P_{2;-1}^{(1,1)}(1,2,3,4,5) &=  A_{2;0}^{(0)}(1,2,3,4,5)\left( -\tfrac{1}{2\eps} \right)
  \label{eq:pole_ttqqg_Ncpm1d23T451_ds1}
\end{align}

\begin{align}
  P_{3;0}^{(1,0)}(1,2,3,4,5) &=
    A_{1;0}^{(0)}(1,2,3,4,5)\left( \tfrac{1}{\eps}\left( L_{m,15}+L_{m,24}-L_{45}-L_{\beta} \right) \right) \nonumber\\&
  + A_{2;0}^{(0)}(1,2,3,4,5)\left( \tfrac{1}{\eps}\left( L_{m,13}+L_{m,25}-L_{35}-L_{\beta} \right) \right) \nonumber\\&
  + A_{3;0}^{(0)}(1,2,3,4,5)\left( -\tfrac{2}{\eps^2} + \tfrac{1}{\eps}\left( L_{45}-L_{35}-L_{\beta} + 2 \right) \right)
  \label{eq:pole_ttqqg_d12T453_ds0} \\
  P_{3;0}^{(1,1)}(1,2,3,4,5) &=
  A_{3;0}^{(0)}(1,2,3,4,5)\left( \tfrac{1}{3\eps} \right)
  \label{eq:pole_ttqqg_d12T453_ds1}
\end{align}

\begin{align}
  P_{3;-2}^{(1,0)}(1,2,3,4,5) &=
  A_{3;0}^{(0)}(1,2,3,4,5)\big( \tfrac{1}{\eps^2} \nonumber\\& \hspace{0.5cm} + \tfrac{1}{\eps}\left( L_{m,14}+L_{m,23}-L_{m,13}-L_{m,24}+L_{34}+L_{\beta} + 2 \big) \right)
  \label{eq:pole_ttqqg_Ncpm2d12T453_ds0} \\
  P_{3;0}^{(1,1)}(1,2,3,4,5) &=
  A_{3;0}^{(0)}(1,2,3,4,5)\left( \tfrac{1}{3\eps} \right)
  \label{eq:pole_ttqqg_Ncpm2d12T453_ds1}
\end{align}

\begin{align}
  P_{4;0}^{(1,0)}(1,2,3,4,5) &=
    A_{1;0}^{(0)}(1,2,3,4,5)\left( \tfrac{1}{\eps}\left( L_{m,13}-L_{m,15}+L_{45}-L_{34} \right) \right) \nonumber\\&
  + A_{2;0}^{(0)}(1,2,3,4,5)\left( \tfrac{1}{\eps}\left( L_{m,24}-L_{m,25}+L_{35}-L_{34} \right) \right) \nonumber\\&
  + A_{4;0}^{(0)}(1,2,3,4,5)\left( -\tfrac{2}{\eps^2} + \tfrac{1}{\eps}\left( -L_{34}-L_{m,15}-L_{m,25} + 2 \right) \right)
  \label{eq:pole_ttqqg_d34T251_ds0} \\
  P_{4;0}^{(1,1)}(1,2,3,4,5) &=
  A_{4;0}^{(0)}(1,2,3,4,5)\left( \tfrac{1}{3\eps} \right)
  \label{eq:pole_ttqqg_d34T251_ds1}
\end{align}

\begin{align}
  P_{4;-2}^{(1,0)}(1,2,3,4,5) &=
  A_{4;0}^{(0)}(1,2,3,4,5)\big( \tfrac{1}{\eps^2} \nonumber\\& \hspace{0.5cm} + \tfrac{1}{\eps}\left( L_{m,14}+L_{m,23}-L_{m,13}-L_{m,24}+L_{34}+L_{\beta} + 2 \big) \right)
  \label{eq:pole_ttqqg_Ncmp2d34T251_ds0} \\
  P_{4;0}^{(1,1)}(1,2,3,4,5) &=
  A_{4;0}^{(0)}(1,2,3,4,5)\left( -\tfrac{1}{2\eps} \right)
  \label{eq:pole_ttqqg_Ncpm2d34T251_ds1}
\end{align}

The closed fermion loops are non-zero but very simple and are given by,

\begin{align}
  P_{I;-1}^{(1,0),N}(1,2,3,4,5) &= 
  A_{I;0}^{(0)}(1,2,3,4,5)\left( -\tfrac{2}{3\eps} \right)
\end{align}

\bibliographystyle{JHEP}
\bibliography{ppttjOe2}

\providecommand{\href}[2]{#2}\begingroup\raggedright\begin{thebibliography}{100}

\bibitem{Dittmaier:2007wz}
S.~Dittmaier, P.~Uwer and S.~Weinzierl, \emph{{NLO QCD corrections to t anti-t
  + jet production at hadron colliders}},
  \href{http://dx.doi.org/10.1103/PhysRevLett.98.262002}{\emph{Phys. Rev.
  Lett.} {\bf 98} (2007) 262002},
  [\href{http://arxiv.org/abs/hep-ph/0703120}{{\tt hep-ph/0703120}}].

\bibitem{Dittmaier:2008uj}
S.~Dittmaier, P.~Uwer and S.~Weinzierl, \emph{{Hadronic top-quark pair
  production in association with a hard jet at next-to-leading order QCD:
  Phenomenological studies for the Tevatron and the LHC}},
  \href{http://dx.doi.org/10.1140/epjc/s10052-008-0816-y}{\emph{Eur. Phys. J.
  C} {\bf 59} (2009) 625--646}, [\href{http://arxiv.org/abs/0810.0452}{{\tt
  0810.0452}}].

\bibitem{Melnikov:2010iu}
K.~Melnikov and M.~Schulze, \emph{{NLO QCD corrections to top quark pair
  production in association with one hard jet at hadron colliders}},
  \href{http://dx.doi.org/10.1016/j.nuclphysb.2010.07.003}{\emph{Nucl. Phys. B}
  {\bf 840} (2010) 129--159}, [\href{http://arxiv.org/abs/1004.3284}{{\tt
  1004.3284}}].

\bibitem{Bevilacqua:2015qha}
G.~Bevilacqua, H.~B. Hartanto, M.~Kraus and M.~Worek, \emph{{Top Quark Pair
  Production in Association with a Jet with Next-to-Leading-Order QCD Off-Shell
  Effects at the Large Hadron Collider}},
  \href{http://dx.doi.org/10.1103/PhysRevLett.116.052003}{\emph{Phys. Rev.
  Lett.} {\bf 116} (2016) 052003}, [\href{http://arxiv.org/abs/1509.09242}{{\tt
  1509.09242}}].

\bibitem{Ossola:2006us}
G.~Ossola, C.~G. Papadopoulos and R.~Pittau, \emph{{Reducing full one-loop
  amplitudes to scalar integrals at the integrand level}},
  \href{http://dx.doi.org/10.1016/j.nuclphysb.2006.11.012}{\emph{Nucl. Phys. B}
  {\bf 763} (2007) 147--169}, [\href{http://arxiv.org/abs/hep-ph/0609007}{{\tt
  hep-ph/0609007}}].

\bibitem{Giele:2008ve}
W.~T. Giele, Z.~Kunszt and K.~Melnikov, \emph{{Full one-loop amplitudes from
  tree amplitudes}},
  \href{http://dx.doi.org/10.1088/1126-6708/2008/04/049}{\emph{JHEP} {\bf 04}
  (2008) 049}, [\href{http://arxiv.org/abs/0801.2237}{{\tt 0801.2237}}].

\bibitem{Berger:2008sj}
C.~F. Berger, Z.~Bern, L.~J. Dixon, F.~Febres~Cordero, D.~Forde, H.~Ita et~al.,
  \emph{{An Automated Implementation of On-Shell Methods for One-Loop
  Amplitudes}}, \href{http://dx.doi.org/10.1103/PhysRevD.78.036003}{\emph{Phys.
  Rev. D} {\bf 78} (2008) 036003}, [\href{http://arxiv.org/abs/0803.4180}{{\tt
  0803.4180}}].

\bibitem{Ellis:2008ir}
R.~K. Ellis, W.~T. Giele, Z.~Kunszt and K.~Melnikov, \emph{{Masses, fermions
  and generalized $D$-dimensional unitarity}},
  \href{http://dx.doi.org/10.1016/j.nuclphysb.2009.07.023}{\emph{Nucl. Phys. B}
  {\bf 822} (2009) 270--282}, [\href{http://arxiv.org/abs/0806.3467}{{\tt
  0806.3467}}].

\bibitem{Bevilacqua:2011xh}
G.~Bevilacqua, M.~Czakon, M.~V. Garzelli, A.~van Hameren, A.~Kardos, C.~G.
  Papadopoulos et~al., \emph{{HELAC-NLO}},
  \href{http://dx.doi.org/10.1016/j.cpc.2012.10.033}{\emph{Comput. Phys.
  Commun.} {\bf 184} (2013) 986--997},
  [\href{http://arxiv.org/abs/1110.1499}{{\tt 1110.1499}}].

\bibitem{Cullen:2011ac}
G.~Cullen, N.~Greiner, G.~Heinrich, G.~Luisoni, P.~Mastrolia, G.~Ossola et~al.,
  \emph{{Automated One-Loop Calculations with GoSam}},
  \href{http://dx.doi.org/10.1140/epjc/s10052-012-1889-1}{\emph{Eur. Phys. J.
  C} {\bf 72} (2012) 1889}, [\href{http://arxiv.org/abs/1111.2034}{{\tt
  1111.2034}}].

\bibitem{Cascioli:2011va}
F.~Cascioli, P.~Maierhofer and S.~Pozzorini, \emph{{Scattering Amplitudes with
  Open Loops}},
  \href{http://dx.doi.org/10.1103/PhysRevLett.108.111601}{\emph{Phys. Rev.
  Lett.} {\bf 108} (2012) 111601}, [\href{http://arxiv.org/abs/1111.5206}{{\tt
  1111.5206}}].

\bibitem{Hoche:2016elu}
S.~H\"oche, P.~Maierh\"ofer, N.~Moretti, S.~Pozzorini and F.~Siegert,
  \emph{{Next-to-leading order QCD predictions for top-quark pair production
  with up to three jets}},
  \href{http://dx.doi.org/10.1140/epjc/s10052-017-4715-y}{\emph{Eur. Phys. J.
  C} {\bf 77} (2017) 145}, [\href{http://arxiv.org/abs/1607.06934}{{\tt
  1607.06934}}].

\bibitem{Alioli:2011as}
S.~Alioli, S.-O. Moch and P.~Uwer, \emph{{Hadronic top-quark pair-production
  with one jet and parton showering}},
  \href{http://dx.doi.org/10.1007/JHEP01(2012)137}{\emph{JHEP} {\bf 01} (2012)
  137}, [\href{http://arxiv.org/abs/1110.5251}{{\tt 1110.5251}}].

\bibitem{Hoeche:2014qda}
S.~Hoeche, F.~Krauss, P.~Maierhoefer, S.~Pozzorini, M.~Schonherr and
  F.~Siegert, \emph{{Next-to-leading order QCD predictions for top-quark pair
  production with up to two jets merged with a parton shower}},
  \href{http://dx.doi.org/10.1016/j.physletb.2015.06.060}{\emph{Phys. Lett. B}
  {\bf 748} (2015) 74--78}, [\href{http://arxiv.org/abs/1402.6293}{{\tt
  1402.6293}}].

\bibitem{Czakon:2015cla}
M.~Czakon, H.~B. Hartanto, M.~Kraus and M.~Worek, \emph{{Matching the
  Nagy-Soper parton shower at next-to-leading order}},
  \href{http://dx.doi.org/10.1007/JHEP06(2015)033}{\emph{JHEP} {\bf 06} (2015)
  033}, [\href{http://arxiv.org/abs/1502.00925}{{\tt 1502.00925}}].

\bibitem{Alioli:2013mxa}
S.~Alioli, P.~Fernandez, J.~Fuster, A.~Irles, S.-O. Moch, P.~Uwer et~al.,
  \emph{{A new observable to measure the top-quark mass at hadron colliders}},
  \href{http://dx.doi.org/10.1140/epjc/s10052-013-2438-2}{\emph{Eur. Phys. J.
  C} {\bf 73} (2013) 2438}, [\href{http://arxiv.org/abs/1303.6415}{{\tt
  1303.6415}}].

\bibitem{Bevilacqua:2017ipv}
G.~Bevilacqua, H.~B. Hartanto, M.~Kraus, M.~Schulze and M.~Worek, \emph{{Top
  quark mass studies with $ t\overline{t}j $ at the LHC}},
  \href{http://dx.doi.org/10.1007/JHEP03(2018)169}{\emph{JHEP} {\bf 03} (2018)
  169}, [\href{http://arxiv.org/abs/1710.07515}{{\tt 1710.07515}}].

\bibitem{Czakon:2010td}
M.~Czakon, \emph{{A novel subtraction scheme for double-real radiation at
  NNLO}}, \href{http://dx.doi.org/10.1016/j.physletb.2010.08.036}{\emph{Phys.
  Lett. B} {\bf 693} (2010) 259--268},
  [\href{http://arxiv.org/abs/1005.0274}{{\tt 1005.0274}}].

\bibitem{Czakon:2013goa}
M.~Czakon, P.~Fiedler and A.~Mitov, \emph{{Total Top-Quark Pair-Production
  Cross Section at Hadron Colliders Through $O(\alpha^4_S)$}},
  \href{http://dx.doi.org/10.1103/PhysRevLett.110.252004}{\emph{Phys. Rev.
  Lett.} {\bf 110} (2013) 252004}, [\href{http://arxiv.org/abs/1303.6254}{{\tt
  1303.6254}}].

\bibitem{Behring:2019iiv}
A.~Behring, M.~Czakon, A.~Mitov, A.~S. Papanastasiou and R.~Poncelet,
  \emph{{Higher order corrections to spin correlations in top quark pair
  production at the LHC}},
  \href{http://dx.doi.org/10.1103/PhysRevLett.123.082001}{\emph{Phys. Rev.
  Lett.} {\bf 123} (2019) 082001}, [\href{http://arxiv.org/abs/1901.05407}{{\tt
  1901.05407}}].

\bibitem{Catani:2019hip}
S.~Catani, S.~Devoto, M.~Grazzini, S.~Kallweit and J.~Mazzitelli,
  \emph{{Top-quark pair production at the LHC: Fully differential QCD
  predictions at NNLO}},
  \href{http://dx.doi.org/10.1007/JHEP07(2019)100}{\emph{JHEP} {\bf 07} (2019)
  100}, [\href{http://arxiv.org/abs/1906.06535}{{\tt 1906.06535}}].

\bibitem{Bonciani:2008az}
R.~Bonciani, A.~Ferroglia, T.~Gehrmann, D.~Maitre and C.~Studerus,
  \emph{{Two-Loop Fermionic Corrections to Heavy-Quark Pair Production: The
  Quark-Antiquark Channel}},
  \href{http://dx.doi.org/10.1088/1126-6708/2008/07/129}{\emph{JHEP} {\bf 07}
  (2008) 129}, [\href{http://arxiv.org/abs/0806.2301}{{\tt 0806.2301}}].

\bibitem{Bonciani:2009nb}
R.~Bonciani, A.~Ferroglia, T.~Gehrmann and C.~Studerus, \emph{{Two-Loop Planar
  Corrections to Heavy-Quark Pair Production in the Quark-Antiquark Channel}},
  \href{http://dx.doi.org/10.1088/1126-6708/2009/08/067}{\emph{JHEP} {\bf 08}
  (2009) 067}, [\href{http://arxiv.org/abs/0906.3671}{{\tt 0906.3671}}].

\bibitem{Bonciani:2010mn}
R.~Bonciani, A.~Ferroglia, T.~Gehrmann, A.~von Manteuffel and C.~Studerus,
  \emph{{Two-Loop Leading Color Corrections to Heavy-Quark Pair Production in
  the Gluon Fusion Channel}},
  \href{http://dx.doi.org/10.1007/JHEP01(2011)102}{\emph{JHEP} {\bf 01} (2011)
  102}, [\href{http://arxiv.org/abs/1011.6661}{{\tt 1011.6661}}].

\bibitem{Bonciani:2013ywa}
R.~Bonciani, A.~Ferroglia, T.~Gehrmann, A.~von Manteuffel and C.~Studerus,
  \emph{{Light-quark two-loop corrections to heavy-quark pair production in the
  gluon fusion channel}},
  \href{http://dx.doi.org/10.1007/JHEP12(2013)038}{\emph{JHEP} {\bf 12} (2013)
  038}, [\href{http://arxiv.org/abs/1309.4450}{{\tt 1309.4450}}].

\bibitem{vonManteuffel:2013uoa}
A.~von Manteuffel and C.~Studerus, \emph{{Massive planar and non-planar double
  box integrals for light Nf contributions to gg-\ensuremath{>}tt}},
  \href{http://dx.doi.org/10.1007/JHEP10(2013)037}{\emph{JHEP} {\bf 10} (2013)
  037}, [\href{http://arxiv.org/abs/1306.3504}{{\tt 1306.3504}}].

\bibitem{DiVita:2018nnh}
S.~Di~Vita, S.~Laporta, P.~Mastrolia, A.~Primo and U.~Schubert, \emph{{Master
  integrals for the NNLO virtual corrections to $\mu e$ scattering in QED: the
  non-planar graphs}},
  \href{http://dx.doi.org/10.1007/JHEP09(2018)016}{\emph{JHEP} {\bf 09} (2018)
  016}, [\href{http://arxiv.org/abs/1806.08241}{{\tt 1806.08241}}].

\bibitem{Mastrolia:2017pfy}
P.~Mastrolia, M.~Passera, A.~Primo and U.~Schubert, \emph{{Master integrals for
  the NNLO virtual corrections to $\mu e$ scattering in QED: the planar
  graphs}}, \href{http://dx.doi.org/10.1007/JHEP11(2017)198}{\emph{JHEP} {\bf
  11} (2017) 198}, [\href{http://arxiv.org/abs/1709.07435}{{\tt 1709.07435}}].

\bibitem{Becchetti:2019tjy}
M.~Becchetti, R.~Bonciani, V.~Casconi, A.~Ferroglia, S.~Lavacca and A.~von
  Manteuffel, \emph{{Master Integrals for the two-loop, non-planar QCD
  corrections to top-quark pair production in the quark-annihilation channel}},
  \href{http://dx.doi.org/10.1007/JHEP08(2019)071}{\emph{JHEP} {\bf 08} (2019)
  071}, [\href{http://arxiv.org/abs/1904.10834}{{\tt 1904.10834}}].

\bibitem{Badger:2021owl}
S.~Badger, E.~Chaubey, H.~B. Hartanto and R.~Marzucca, \emph{{Two-loop leading
  colour QCD helicity amplitudes for top quark pair production in the gluon
  fusion channel}},
  \href{http://dx.doi.org/10.1007/JHEP06(2021)163}{\emph{JHEP} {\bf 06} (2021)
  163}, [\href{http://arxiv.org/abs/2102.13450}{{\tt 2102.13450}}].

\bibitem{Badger:2011yu}
S.~Badger, R.~Sattler and V.~Yundin, \emph{{One-Loop Helicity Amplitudes for
  $t\bar{t}$ Production at Hadron Colliders}},
  \href{http://dx.doi.org/10.1103/PhysRevD.83.074020}{\emph{Phys. Rev. D} {\bf
  83} (2011) 074020}, [\href{http://arxiv.org/abs/1101.5947}{{\tt 1101.5947}}].

\bibitem{Syrrakos:2021nij}
N.~Syrrakos, \emph{{One-loop Feynman integrals for 2 \textrightarrow{} 3
  scattering involving many scales including internal masses}},
  \href{http://dx.doi.org/10.1007/JHEP10(2021)041}{\emph{JHEP} {\bf 10} (2021)
  041}, [\href{http://arxiv.org/abs/2107.02106}{{\tt 2107.02106}}].

\bibitem{Francesco:2019yqt}
F.~Moriello, \emph{{Generalised power series expansions for the elliptic planar
  families of Higgs + jet production at two loops}},
  \href{http://dx.doi.org/10.1007/JHEP01(2020)150}{\emph{JHEP} {\bf 01} (2020)
  150}, [\href{http://arxiv.org/abs/1907.13234}{{\tt 1907.13234}}].

\bibitem{Hidding:2020ytt}
M.~Hidding, \emph{{DiffExp, a Mathematica package for computing Feynman
  integrals in terms of one-dimensional series expansions}},
  \href{http://arxiv.org/abs/2006.05510}{{\tt 2006.05510}}.

\bibitem{Abreu:2020jxa}
S.~Abreu, H.~Ita, F.~Moriello, B.~Page, W.~Tschernow and M.~Zeng,
  \emph{{Two-Loop Integrals for Planar Five-Point One-Mass Processes}},
  \href{http://dx.doi.org/10.1007/JHEP11(2020)117}{\emph{JHEP} {\bf 11} (2020)
  117}, [\href{http://arxiv.org/abs/2005.04195}{{\tt 2005.04195}}].

\bibitem{Becchetti:2020wof}
M.~Becchetti, R.~Bonciani, V.~Del~Duca, V.~Hirschi, F.~Moriello and
  A.~Schweitzer, \emph{{Next-to-leading order corrections to light-quark mixed
  QCD-EW contributions to Higgs boson production}},
  \href{http://dx.doi.org/10.1103/PhysRevD.103.054037}{\emph{Phys. Rev. D} {\bf
  103} (2021) 054037}, [\href{http://arxiv.org/abs/2010.09451}{{\tt
  2010.09451}}].

\bibitem{Bonciani:2021zzf}
R.~Bonciani, L.~Buonocore, M.~Grazzini, S.~Kallweit, N.~Rana, F.~Tramontano
  et~al., \emph{{Mixed Strong-Electroweak Corrections to the Drell-Yan
  Process}},
  \href{http://dx.doi.org/10.1103/PhysRevLett.128.012002}{\emph{Phys. Rev.
  Lett.} {\bf 128} (2022) 012002}, [\href{http://arxiv.org/abs/2106.11953}{{\tt
  2106.11953}}].

\bibitem{Armadillo:2022bgm}
T.~Armadillo, R.~Bonciani, S.~Devoto, N.~Rana and A.~Vicini, \emph{{Two-loop
  mixed QCD-EW corrections to neutral current Drell-Yan}},
  \href{http://arxiv.org/abs/2201.01754}{{\tt 2201.01754}}.

\bibitem{Lee:2017qql}
R.~N. Lee, A.~V. Smirnov and V.~A. Smirnov, \emph{{Solving differential
  equations for Feynman integrals by expansions near singular points}},
  \href{http://dx.doi.org/10.1007/JHEP03(2018)008}{\emph{JHEP} {\bf 03} (2018)
  008}, [\href{http://arxiv.org/abs/1709.07525}{{\tt 1709.07525}}].

\bibitem{Mandal:2018cdj}
M.~K. Mandal and X.~Zhao, \emph{{Evaluating multi-loop Feynman integrals
  numerically through differential equations}},
  \href{http://dx.doi.org/10.1007/JHEP03(2019)190}{\emph{JHEP} {\bf 03} (2019)
  190}, [\href{http://arxiv.org/abs/1812.03060}{{\tt 1812.03060}}].

\bibitem{Liu:2021wks}
X.~Liu and Y.-Q. Ma, \emph{{Multiloop corrections for collider processes using
  auxiliary mass flow}},  \href{http://arxiv.org/abs/2107.01864}{{\tt
  2107.01864}}.

\bibitem{Liu:2022chg}
X.~Liu and Y.-Q. Ma, \emph{{AMFlow: a Mathematica Package for Feynman integrals
  computation via Auxiliary Mass Flow}},
  \href{http://arxiv.org/abs/2201.11669}{{\tt 2201.11669}}.

\bibitem{Liu:2022tji}
Z.-F. Liu and Y.-Q. Ma, \emph{{Automatic computation of Feynman integrals
  containing linear propagators via auxiliary mass flow}},
  \href{http://arxiv.org/abs/2201.11636}{{\tt 2201.11636}}.

\bibitem{Gehrmann:2015bfy}
T.~Gehrmann, J.~Henn and N.~Lo~Presti, \emph{{Analytic form of the two-loop
  planar five-gluon all-plus-helicity amplitude in QCD}},
  \href{http://dx.doi.org/10.1103/PhysRevLett.116.062001}{\emph{Phys. Rev.
  Lett.} {\bf 116} (2016) 062001}, [\href{http://arxiv.org/abs/1511.05409}{{\tt
  1511.05409}}].

\bibitem{Badger:2018enw}
S.~Badger, C.~Br\o{}nnum-Hansen, H.~B. Hartanto and T.~Peraro, \emph{{Analytic
  helicity amplitudes for two-loop five-gluon scattering: the single-minus
  case}}, \href{http://dx.doi.org/10.1007/JHEP01(2019)186}{\emph{JHEP} {\bf 01}
  (2019) 186}, [\href{http://arxiv.org/abs/1811.11699}{{\tt 1811.11699}}].

\bibitem{Abreu:2018aqd}
S.~Abreu, L.~J. Dixon, E.~Herrmann, B.~Page and M.~Zeng, \emph{{The two-loop
  five-point amplitude in $\mathcal{N} =4$ super-Yang-Mills theory}},
  \href{http://dx.doi.org/10.1103/PhysRevLett.122.121603}{\emph{Phys. Rev.
  Lett.} {\bf 122} (2019) 121603}, [\href{http://arxiv.org/abs/1812.08941}{{\tt
  1812.08941}}].

\bibitem{Chicherin:2018yne}
D.~Chicherin, T.~Gehrmann, J.~Henn, P.~Wasser, Y.~Zhang and S.~Zoia,
  \emph{{Analytic result for a two-loop five-particle amplitude}},
  \href{http://dx.doi.org/10.1103/PhysRevLett.122.121602}{\emph{Phys. Rev.
  Lett.} {\bf 122} (2019) 121602}, [\href{http://arxiv.org/abs/1812.11057}{{\tt
  1812.11057}}].

\bibitem{Chicherin:2019xeg}
D.~Chicherin, T.~Gehrmann, J.~M. Henn, P.~Wasser, Y.~Zhang and S.~Zoia,
  \emph{{The two-loop five-particle amplitude in $ \mathcal{N} $ = 8
  supergravity}}, \href{http://dx.doi.org/10.1007/JHEP03(2019)115}{\emph{JHEP}
  {\bf 03} (2019) 115}, [\href{http://arxiv.org/abs/1901.05932}{{\tt
  1901.05932}}].

\bibitem{Abreu:2019rpt}
S.~Abreu, L.~J. Dixon, E.~Herrmann, B.~Page and M.~Zeng, \emph{{The two-loop
  five-point amplitude in $ \mathcal{N} $ = 8 supergravity}},
  \href{http://dx.doi.org/10.1007/JHEP03(2019)123}{\emph{JHEP} {\bf 03} (2019)
  123}, [\href{http://arxiv.org/abs/1901.08563}{{\tt 1901.08563}}].

\bibitem{Abreu:2018zmy}
S.~Abreu, J.~Dormans, F.~Febres~Cordero, H.~Ita and B.~Page, \emph{{Analytic
  Form of Planar Two-Loop Five-Gluon Scattering Amplitudes in QCD}},
  \href{http://dx.doi.org/10.1103/PhysRevLett.122.082002}{\emph{Phys. Rev.
  Lett.} {\bf 122} (2019) 082002}, [\href{http://arxiv.org/abs/1812.04586}{{\tt
  1812.04586}}].

\bibitem{Abreu:2019odu}
S.~Abreu, J.~Dormans, F.~Febres~Cordero, H.~Ita, B.~Page and V.~Sotnikov,
  \emph{{Analytic Form of the Planar Two-Loop Five-Parton Scattering Amplitudes
  in QCD}}, \href{http://dx.doi.org/10.1007/JHEP05(2019)084}{\emph{JHEP} {\bf
  05} (2019) 084}, [\href{http://arxiv.org/abs/1904.00945}{{\tt 1904.00945}}].

\bibitem{Badger:2019djh}
S.~Badger, D.~Chicherin, T.~Gehrmann, G.~Heinrich, J.~Henn, T.~Peraro et~al.,
  \emph{{Analytic form of the full two-loop five-gluon all-plus helicity
  amplitude}},
  \href{http://dx.doi.org/10.1103/PhysRevLett.123.071601}{\emph{Phys. Rev.
  Lett.} {\bf 123} (2019) 071601}, [\href{http://arxiv.org/abs/1905.03733}{{\tt
  1905.03733}}].

\bibitem{Abreu:2020cwb}
S.~Abreu, B.~Page, E.~Pascual and V.~Sotnikov, \emph{{Leading-Color Two-Loop
  QCD Corrections for Three-Photon Production at Hadron Colliders}},
  \href{http://dx.doi.org/10.1007/JHEP01(2021)078}{\emph{JHEP} {\bf 01} (2021)
  078}, [\href{http://arxiv.org/abs/2010.15834}{{\tt 2010.15834}}].

\bibitem{Chawdhry:2020for}
H.~A. Chawdhry, M.~Czakon, A.~Mitov and R.~Poncelet, \emph{{Two-loop
  leading-color helicity amplitudes for three-photon production at the LHC}},
  \href{http://dx.doi.org/10.1007/JHEP06(2021)150}{\emph{JHEP} {\bf 06} (2021)
  150}, [\href{http://arxiv.org/abs/2012.13553}{{\tt 2012.13553}}].

\bibitem{Caron-Huot:2020vlo}
S.~Caron-Huot, D.~Chicherin, J.~Henn, Y.~Zhang and S.~Zoia, \emph{{Multi-Regge
  Limit of the Two-Loop Five-Point Amplitudes in $\mathcal{N} = 4$ Super
  Yang-Mills and $\mathcal{N} = 8$ Supergravity}},
  \href{http://dx.doi.org/10.1007/JHEP10(2020)188}{\emph{JHEP} {\bf 10} (2020)
  188}, [\href{http://arxiv.org/abs/2003.03120}{{\tt 2003.03120}}].

\bibitem{DeLaurentis:2020qle}
G.~De~Laurentis and D.~Ma\^\i{}tre, \emph{{Two-Loop Five-Parton Leading-Colour
  Finite Remainders in the Spinor-Helicity Formalism}},
  \href{http://dx.doi.org/10.1007/JHEP02(2021)016}{\emph{JHEP} {\bf 02} (2021)
  016}, [\href{http://arxiv.org/abs/2010.14525}{{\tt 2010.14525}}].

\bibitem{Agarwal:2021grm}
B.~Agarwal, F.~Buccioni, A.~von Manteuffel and L.~Tancredi, \emph{{Two-loop
  leading colour QCD corrections to $q \bar{q} \to \gamma \gamma g$ and $q g
  \to \gamma \gamma q$}},
  \href{http://dx.doi.org/10.1007/JHEP04(2021)201}{\emph{JHEP} {\bf 04} (2021)
  201}, [\href{http://arxiv.org/abs/2102.01820}{{\tt 2102.01820}}].

\bibitem{Abreu:2021oya}
S.~Abreu, F.~F. Cordero, H.~Ita, B.~Page and V.~Sotnikov, \emph{{Leading-color
  two-loop QCD corrections for three-jet production at hadron colliders}},
  \href{http://dx.doi.org/10.1007/JHEP07(2021)095}{\emph{JHEP} {\bf 07} (2021)
  095}, [\href{http://arxiv.org/abs/2102.13609}{{\tt 2102.13609}}].

\bibitem{Agarwal:2021vdh}
B.~Agarwal, F.~Buccioni, A.~von Manteuffel and L.~Tancredi, \emph{{Two-Loop
  Helicity Amplitudes for Diphoton Plus Jet Production in Full Color}},
  \href{http://dx.doi.org/10.1103/PhysRevLett.127.262001}{\emph{Phys. Rev.
  Lett.} {\bf 127} (2021) 262001}, [\href{http://arxiv.org/abs/2105.04585}{{\tt
  2105.04585}}].

\bibitem{Chawdhry:2021mkw}
H.~A. Chawdhry, M.~Czakon, A.~Mitov and R.~Poncelet, \emph{{Two-loop
  leading-colour QCD helicity amplitudes for two-photon plus jet production at
  the LHC}}, \href{http://dx.doi.org/10.1007/JHEP07(2021)164}{\emph{JHEP} {\bf
  07} (2021) 164}, [\href{http://arxiv.org/abs/2103.04319}{{\tt 2103.04319}}].

\bibitem{Badger:2021nhg}
S.~Badger, H.~B. Hartanto and S.~Zoia, \emph{{Two-Loop QCD Corrections to
  $Wb\bar{b}$ Production at Hadron Colliders}},
  \href{http://dx.doi.org/10.1103/PhysRevLett.127.012001}{\emph{Phys. Rev.
  Lett.} {\bf 127} (2021) 012001}, [\href{http://arxiv.org/abs/2102.02516}{{\tt
  2102.02516}}].

\bibitem{Badger:2021imn}
S.~Badger, C.~Br\o{}nnum-Hansen, D.~Chicherin, T.~Gehrmann, H.~B. Hartanto,
  J.~Henn et~al., \emph{{Virtual QCD corrections to gluon-initiated diphoton
  plus jet production at hadron colliders}},
  \href{http://dx.doi.org/10.1007/JHEP11(2021)083}{\emph{JHEP} {\bf 11} (2021)
  083}, [\href{http://arxiv.org/abs/2106.08664}{{\tt 2106.08664}}].

\bibitem{Abreu:2021asb}
S.~Abreu, F.~F. Cordero, H.~Ita, M.~Klinkert, B.~Page and V.~Sotnikov,
  \emph{{Leading-Color Two-Loop Amplitudes for Four Partons and a W Boson in
  QCD}},  \href{http://arxiv.org/abs/2110.07541}{{\tt 2110.07541}}.

\bibitem{Badger:2021ega}
S.~Badger, H.~B. Hartanto, J.~Kry\'s and S.~Zoia, \emph{{Two-loop
  leading-colour QCD helicity amplitudes for Higgs boson production in
  association with a bottom-quark pair at the LHC}},
  \href{http://dx.doi.org/10.1007/JHEP11(2021)012}{\emph{JHEP} {\bf 11} (2021)
  012}, [\href{http://arxiv.org/abs/2107.14733}{{\tt 2107.14733}}].

\bibitem{Badger:2022ncb}
S.~Badger, H.~B. Hartanto, J.~Kry\'s and S.~Zoia, \emph{{Two-loop leading
  colour helicity amplitudes for $W^\pm\gamma+j$ production at the LHC}},
  \href{http://arxiv.org/abs/2201.04075}{{\tt 2201.04075}}.

\bibitem{Peraro:2016wsq}
T.~Peraro, \emph{{Scattering amplitudes over finite fields and multivariate
  functional reconstruction}},
  \href{http://dx.doi.org/10.1007/JHEP12(2016)030}{\emph{JHEP} {\bf 12} (2016)
  030}, [\href{http://arxiv.org/abs/1608.01902}{{\tt 1608.01902}}].

\bibitem{Peraro:2019svx}
T.~Peraro, \emph{{FiniteFlow: multivariate functional reconstruction using
  finite fields and dataflow graphs}},
  \href{http://dx.doi.org/10.1007/JHEP07(2019)031}{\emph{JHEP} {\bf 07} (2019)
  031}, [\href{http://arxiv.org/abs/1905.08019}{{\tt 1905.08019}}].

\bibitem{Hodges:2009hk}
A.~Hodges, \emph{{Eliminating spurious poles from gauge-theoretic amplitudes}},
  \href{http://dx.doi.org/10.1007/JHEP05(2013)135}{\emph{JHEP} {\bf 05} (2013)
  135}, [\href{http://arxiv.org/abs/0905.1473}{{\tt 0905.1473}}].

\bibitem{Bern:1994fz}
Z.~Bern, L.~J. Dixon and D.~A. Kosower, \emph{{One loop corrections to two
  quark three gluon amplitudes}},
  \href{http://dx.doi.org/10.1016/0550-3213(94)00542-M}{\emph{Nucl. Phys. B}
  {\bf 437} (1995) 259--304}, [\href{http://arxiv.org/abs/hep-ph/9409393}{{\tt
  hep-ph/9409393}}].

\bibitem{Kunszt:1994nq}
Z.~Kunszt, A.~Signer and Z.~Trocsanyi, \emph{{One loop radiative corrections to
  the helicity amplitudes of QCD processes involving four quarks and one
  gluon}}, \href{http://dx.doi.org/10.1016/0370-2693(94)90568-1}{\emph{Phys.
  Lett. B} {\bf 336} (1994) 529--536},
  [\href{http://arxiv.org/abs/hep-ph/9405386}{{\tt hep-ph/9405386}}].

\bibitem{catani2001one}
S.~Catani, S.~Dittmaier and Z.~Trocsanyi, \emph{One-loop singular behaviour of
  qcd and susy qcd amplitudes with massive partons}, {\emph{Physics Letters B}
  {\bf 500} (2001) 149--160}.

\bibitem{Catani:1996vz}
S.~Catani and M.~H. Seymour, \emph{{A General algorithm for calculating jet
  cross-sections in NLO QCD}},
  \href{http://dx.doi.org/10.1016/S0550-3213(96)00589-5}{\emph{Nucl. Phys. B}
  {\bf 485} (1997) 291--419}, [\href{http://arxiv.org/abs/hep-ph/9605323}{{\tt
  hep-ph/9605323}}].

\bibitem{Kleiss:1985yh}
R.~Kleiss and W.~J. Stirling, \emph{{Spinor Techniques for Calculating p anti-p
  ---\ensuremath{>} W+- / Z0 + Jets}},
  \href{http://dx.doi.org/10.1016/0550-3213(85)90285-8}{\emph{Nucl. Phys. B}
  {\bf 262} (1985) 235--262}.

\bibitem{Campbell:2012uf}
J.~M. Campbell and R.~K. Ellis, \emph{{Top-Quark Processes at NLO in Production
  and Decay}}, \href{http://dx.doi.org/10.1088/0954-3899/42/1/015005}{\emph{J.
  Phys. G} {\bf 42} (2015) 015005}, [\href{http://arxiv.org/abs/1204.1513}{{\tt
  1204.1513}}].

\bibitem{Badger:2016uuq}
S.~Badger, \emph{{Automating QCD amplitudes with on-shell methods}},
  \href{http://dx.doi.org/10.1088/1742-6596/762/1/012057}{\emph{J. Phys. Conf.
  Ser.} {\bf 762} (2016) 012057}, [\href{http://arxiv.org/abs/1605.02172}{{\tt
  1605.02172}}].

\bibitem{Buciuni:Thesis}
F.~Buciuni, \emph{Applications of Modern Methods for Scattering Amplitudes}.
\newblock PhD thesis, Department of Physics, Durham University, 2018.

\bibitem{Pogel:2021buy}
{P\"{o}gel, Sebastian}, \emph{Unitarity Approaches to Two-Loop All-Plus
  Amplitudes}.
\newblock PhD thesis, IPhT Scalay, 2021.

\bibitem{Nogueira:1991ex}
P.~Nogueira, \emph{{Automatic Feynman graph generation}},
  \href{http://dx.doi.org/10.1006/jcph.1993.1074}{\emph{J. Comput. Phys.} {\bf
  105} (1993) 279--289}.

\bibitem{Kuipers:2012rf}
J.~Kuipers, T.~Ueda, J.~A.~M. Vermaseren and J.~Vollinga, \emph{{FORM version
  4.0}}, \href{http://dx.doi.org/10.1016/j.cpc.2012.12.028}{\emph{Comput. Phys.
  Commun.} {\bf 184} (2013) 1453--1467},
  [\href{http://arxiv.org/abs/1203.6543}{{\tt 1203.6543}}].

\bibitem{Ruijl:2017dtg}
B.~Ruijl, T.~Ueda and J.~Vermaseren, \emph{{FORM version 4.2}},
  \href{http://arxiv.org/abs/1707.06453}{{\tt 1707.06453}}.

\bibitem{Cullen:2010jv}
G.~Cullen, M.~Koch-Janusz and T.~Reiter, \emph{{Spinney: A Form Library for
  Helicity Spinors}},
  \href{http://dx.doi.org/10.1016/j.cpc.2011.06.007}{\emph{Comput. Phys.
  Commun.} {\bf 182} (2011) 2368--2387},
  [\href{http://arxiv.org/abs/1008.0803}{{\tt 1008.0803}}].

\bibitem{Mastrolia:2016dhn}
P.~Mastrolia, T.~Peraro and A.~Primo, \emph{{Adaptive Integrand Decomposition
  in parallel and orthogonal space}},
  \href{http://dx.doi.org/10.1007/JHEP08(2016)164}{\emph{JHEP} {\bf 08} (2016)
  164}, [\href{http://arxiv.org/abs/1605.03157}{{\tt 1605.03157}}].

\bibitem{Tkachov:1981wb}
F.~V. Tkachov, \emph{{A Theorem on Analytical Calculability of Four Loop
  Renormalization Group Functions}},
  \href{http://dx.doi.org/10.1016/0370-2693(81)90288-4}{\emph{Phys. Lett.} {\bf
  100B} (1981) 65--68}.

\bibitem{Chetyrkin:1981qh}
K.~G. Chetyrkin and F.~V. Tkachov, \emph{{Integration by Parts: The Algorithm
  to Calculate beta Functions in 4 Loops}},
  \href{http://dx.doi.org/10.1016/0550-3213(81)90199-1}{\emph{Nucl. Phys. B}
  {\bf 192} (1981) 159--204}.

\bibitem{Laporta:2001dd}
S.~Laporta, \emph{{High precision calculation of multiloop Feynman integrals by
  difference equations}},
  \href{http://dx.doi.org/10.1016/S0217-751X(00)00215-7,
  10.1142/S0217751X00002157}{\emph{Int. J. Mod. Phys.} {\bf A15} (2000)
  5087--5159}, [\href{http://arxiv.org/abs/hep-ph/0102033}{{\tt
  hep-ph/0102033}}].

\bibitem{Lee:2012cn}
R.~N. Lee, \emph{{Presenting LiteRed: a tool for the Loop InTEgrals
  REDuction}},  \href{http://arxiv.org/abs/1212.2685}{{\tt 1212.2685}}.

\bibitem{Britto:2011cr}
R.~Britto and E.~Mirabella, \emph{{External leg corrections in the unitarity
  method}}, \href{http://dx.doi.org/10.1007/JHEP01(2012)045}{\emph{JHEP} {\bf
  01} (2012) 045}, [\href{http://arxiv.org/abs/1109.5106}{{\tt 1109.5106}}].

\bibitem{Badger:2017gta}
S.~Badger, C.~Br\o{}nnum-Hansen, F.~Buciuni and D.~O'Connell, \emph{{A
  unitarity compatible approach to one-loop amplitudes with massive fermions}},
  \href{http://dx.doi.org/10.1007/JHEP06(2017)141}{\emph{JHEP} {\bf 06} (2017)
  141}, [\href{http://arxiv.org/abs/1703.05734}{{\tt 1703.05734}}].

\bibitem{Chetyrkin:1979bj}
K.~G. Chetyrkin, A.~L. Kataev and F.~V. Tkachov, \emph{{Higher Order
  Corrections to Sigma-t (e+ e- ---\ensuremath{>} Hadrons) in Quantum
  Chromodynamics}},
  \href{http://dx.doi.org/10.1016/0370-2693(79)90596-3}{\emph{Phys. Lett. B}
  {\bf 85} (1979) 277--279}.

\bibitem{Lee:2013mka}
R.~N. Lee, \emph{{LiteRed 1.4: a powerful tool for reduction of multiloop
  integrals}}, \href{http://dx.doi.org/10.1088/1742-6596/523/1/012059}{\emph{J.
  Phys. Conf. Ser.} {\bf 523} (2014) 012059},
  [\href{http://arxiv.org/abs/1310.1145}{{\tt 1310.1145}}].

\bibitem{Kotikov:1990kg}
A.~V. Kotikov, \emph{{Differential equations method: New technique for massive
  Feynman diagrams calculation}},
  \href{http://dx.doi.org/10.1016/0370-2693(91)90413-K}{\emph{Phys. Lett. B}
  {\bf 254} (1991) 158--164}.

\bibitem{Remiddi:1997ny}
E.~Remiddi, \emph{{Differential equations for Feynman graph amplitudes}},
  {\emph{Nuovo Cim. A} {\bf 110} (1997) 1435--1452},
  [\href{http://arxiv.org/abs/hep-th/9711188}{{\tt hep-th/9711188}}].

\bibitem{Gehrmann:1999as}
T.~Gehrmann and E.~Remiddi, \emph{{Differential equations for two loop four
  point functions}},
  \href{http://dx.doi.org/10.1016/S0550-3213(00)00223-6}{\emph{Nucl. Phys. B}
  {\bf 580} (2000) 485--518}, [\href{http://arxiv.org/abs/hep-ph/9912329}{{\tt
  hep-ph/9912329}}].

\bibitem{Henn:2013pwa}
J.~M. Henn, \emph{{Multiloop integrals in dimensional regularization made
  simple}}, \href{http://dx.doi.org/10.1103/PhysRevLett.110.251601}{\emph{Phys.
  Rev. Lett.} {\bf 110} (2013) 251601},
  [\href{http://arxiv.org/abs/1304.1806}{{\tt 1304.1806}}].

\bibitem{Goncharov:1998kja}
A.~B. Goncharov, \emph{{Multiple polylogarithms, cyclotomy and modular
  complexes}}, \href{http://dx.doi.org/10.4310/MRL.1998.v5.n4.a7}{\emph{Math.
  Res. Lett.} {\bf 5} (1998) 497--516},
  [\href{http://arxiv.org/abs/1105.2076}{{\tt 1105.2076}}].

\bibitem{Goncharov:2001iea}
A.~B. Goncharov, \emph{{Multiple polylogarithms and mixed Tate motives}},
  \href{http://arxiv.org/abs/math/0103059}{{\tt math/0103059}}.

\bibitem{Panzer:2014caa}
E.~Panzer, \emph{{Algorithms for the symbolic integration of hyperlogarithms
  with applications to Feynman integrals}},
  \href{http://dx.doi.org/10.1016/j.cpc.2014.10.019}{\emph{Comput. Phys.
  Commun.} {\bf 188} (2015) 148--166},
  [\href{http://arxiv.org/abs/1403.3385}{{\tt 1403.3385}}].

\bibitem{Duhr:2019tlz}
C.~Duhr and F.~Dulat, \emph{{PolyLogTools \textemdash{} polylogs for the
  masses}}, \href{http://dx.doi.org/10.1007/JHEP08(2019)135}{\emph{JHEP} {\bf
  08} (2019) 135}, [\href{http://arxiv.org/abs/1904.07279}{{\tt 1904.07279}}].

\bibitem{Borowka:2017idc}
S.~Borowka, G.~Heinrich, S.~Jahn, S.~P. Jones, M.~Kerner, J.~Schlenk et~al.,
  \emph{{pySecDec: a toolbox for the numerical evaluation of multi-scale
  integrals}}, \href{http://dx.doi.org/10.1016/j.cpc.2017.09.015}{\emph{Comput.
  Phys. Commun.} {\bf 222} (2018) 313--326},
  [\href{http://arxiv.org/abs/1703.09692}{{\tt 1703.09692}}].

\bibitem{Dubovyk:2022frj}
I.~Dubovyk, A.~Freitas, J.~Gluza, K.~Grzanka, M.~Hidding and J.~Usovitsch,
  \emph{{Evaluation of multi-loop multi-scale Feynman integrals for precision
  physics}},  \href{http://arxiv.org/abs/2201.02576}{{\tt 2201.02576}}.

\bibitem{Heller:2021qkz}
M.~Heller and A.~von Manteuffel, \emph{{MultivariateApart: Generalized partial
  fractions}}, \href{http://dx.doi.org/10.1016/j.cpc.2021.108174}{\emph{Comput.
  Phys. Commun.} {\bf 271} (2022) 108174},
  [\href{http://arxiv.org/abs/2101.08283}{{\tt 2101.08283}}].

\end{thebibliography}\endgroup

\end{document}